\RequirePackage{amsthm}

\documentclass{article}
\usepackage{geometry}
\newtheorem{observation}{Observation}
\newtheorem{corollary}{Corollary}
\newtheorem{lemma}{Lemma}
\usepackage{graphicx}%
\usepackage{multirow}%
\usepackage{amsmath,amssymb,amsfonts}%
\usepackage{amsthm}%
\usepackage{mathrsfs}%
\usepackage[title]{appendix}%
\usepackage{xcolor}%
\usepackage{textcomp}%
\usepackage{manyfoot}%
\usepackage{booktabs}%
\usepackage{algorithm}%
\usepackage{algorithmicx}%
\usepackage{algpseudocode}%
\usepackage{listings}%

\theoremstyle{thmstyleone}%
\newtheorem{theorem}{Theorem}%

\theoremstyle{thmstyletwo}%

\theoremstyle{thmstylethree}%

\begin{document}

\title{Convexity of near-optimal orthogonal-pair-free sets on the unit sphere}
\author{Apurva Mudgal, Department of Computer Science and Engineering,\\
Indian Institute of Technology Ropar,\\
Rupnagar, 140001, Punjab, India }
\maketitle

\abstract{
 A subset $S$ of the unit sphere $\mathbb{S}^2$ is called orthogonal-pair-free if and only if there
   do not exist two distinct points $u, v \in S$ at distance $\frac{\pi}{2}$ from each other.  Witsenhausen \cite{witsenhausen} asked the following question:\\

   {\it What is the least upper bound $\alpha_3$ on the Lesbegue measure of any measurable orthogonal-pair-free subset of $\mathbb{S}^2$?}\\
   
   We prove the following result in this paper:
   Let $\mathcal{A}$ be the collection of all orthogonal-pair-free sets $S$ such that $S$ consists of a finite number of mutually disjoint convex sets. Then, $\alpha_3 = \limsup_{S \in \mathcal{A}} \mu(S)$.
   
   Thus, if the double cap conjecture \cite{kalai1} is not true, there is a set in $\mathcal{A}$ with measure strictly greater than the measure of the double cap.
}

\maketitle

\section{Introduction}

Let $\mathbb{S}^2$ be the surface of the sphere of unit radius in $\mathbb{R}^3$, with center at the origin $o$. Two points $p,q \in \mathbb{S}^2$ form an {\it orthogonal pair} if and only if the angle between the rays $\overrightarrow{op}$ and $\overrightarrow{oq}$ is $\frac{\pi}{2}$.
We call a set $A \subset \mathbb{S}^2$ {\it orthogonal-pair-free} if and only if $A$ has no orthogonal pairs.

For a measurable set $S$ on the sphere, let $\mu(S)$ denote its measure (see Section \ref{prelim-measure}; we assume that $\mu(\mathbb{S}^2)=1$). Witsenhausen \cite{witsenhausen} raised the question of computing $\alpha_3 = \limsup_{S} \mu(S)$, where $S$ varies over all measurable orthogonal-pair-free sets on the sphere.
The subscript in $\alpha_3$ refers to the dimension;
for general dimension $d$, we use $\alpha_d$ to denote $\limsup \mu(S)$, where $S$ is a measurable orthogonal-pair-free set of the unit sphere $\mathbb{S}^d$ in $\mathbb{R}^d$ and $\mu$ is the Lebesgue measure on $\mathbb{S}^{d-1}$.

Frankl and Wilson \cite{frankl-wilson} prove the asymptotic upper bound that $\alpha_d \leq  (1+o(1)) \cdot (1.139 \ldots)^{-d}$. Raigorodskii \cite{raigorodskii} improved this bound to $\alpha_d \leq (1+o(1)) \cdot (1.154 \ldots)^{-d}$.

Gil Kalai \cite{kalai1} conjectured that the above limit superior is achieved by the {\it double cap} i.e., the union of the interiors of the two circles of radius $\frac{\pi}{4}$ with centers at the north and south poles.
This conjecture is known as the {\it double cap conjecture} in literature. If the double cap conjecture is true, it implies that $\alpha_d = (\sqrt{2})^{-d} = (1.414 \ldots)^{-d}$.

Witsenhausen \cite{witsenhausen} proved an upper bound of $\frac{1}{3}$ on $\alpha_3$ (for dimension $d > 3$, the bound proved for $\alpha_d$ is $\frac{1}{d}$). DeCorte and Pikhurko \cite{decorte1} improved the upper bound on $\alpha_3$ to $0.313$. They further prove using harmonic analysis, that for every dimension $d \geq 3$, there is a measurable orthogonal-pair-free set of maximum measure i.e., of measure exactly $\alpha_d$.

The bound on $\alpha_3$ has been improved further to $0.308$ (Zhao, \cite{complete-positivity}), $0.30153$ \cite{complete-positivity}, and recently to $0.297742$ \cite{distance-avoiding}. Further, improved bounds have been established for dimensions up to $d=8$ \cite{complete-positivity, distance-avoiding}.\\

{\it Our results.} We say that two sets $A, B$ are {\it mutually disjoint} if and only if $cl(A) \cap cl(B) = \phi$. Let $\mathcal{A}$ be the set of all orthogonal-pair-free subsets $S$ of the unit sphere $\mathbb{S}^2$ such that $S$ consists of a {\it finite number of mutually disjoint convex sets}. As every convex set is measurable, we conclude that all sets in $\mathcal{A}$ are measurable. The main result of the this paper is the following:

\begin{theorem}
\label{main-theorem}
If the double cap conjecture is not true, there exists a set $S \in \mathcal{A}$ with $\mu(S) > \frac{1}{\sqrt{2}}$.
\end{theorem}

Thus, it suffices to look for counterexamples to the double cap conjecture in the set $\mathcal{A}$. In fact, we establish Theorem \ref{main-theorem} by proving that for every measurable orthogonal-pair-free set $S$ and every $\epsilon > 0$, there exists a set $S' \in \mathcal{A}$ such that $\mu(S') > \mu(S) - \epsilon$. Equivalently, we show that $\beta_3 = \alpha_3$, where $\beta_3=\limsup_{S \in \mathcal{A}} \mu(S)$.

Let $\mathcal{A}_k \subset \mathcal{A}$ be the collection of all sets which consist of at most $k$ mutually disjoint convex sets. We also prove the following theorem:\\

\begin{theorem}
\label{finite-k}
For every $k \in \mathbb{N}$, there is a set $M^*_k \in \mathcal{A}_k$
such that $\mu(M^*_k) = \limsup_{S \in \mathcal{A}_k} \mu(S)$.
\end{theorem}

We conclude the paper by observing that the double cap conjecture is equivalent to showing that $M^*_k$ is the double cap for every $k \in \mathbb{N}$.\\

{\it Our results.} The proof of Theorem \ref{main-theorem} spans Sections $2$ and $3$. The proof consists of two main steps. Let $\mathcal{B}$ be the set of all orthogonal-pair-free subsets $S$ of the unit sphere such that $S$ consists of a finite number of almost disjoint dyadic cells (see Section \ref{dyadic-decompose}). In Section $2$, we show that $\alpha_3 = \limsup_{S \in \mathcal{B}} \mu(S)$. In Section $3$, we define the {\it convexification operation} $conv( \cdot )$. For any set $S \in \mathcal{B}$, $conv(S) \in \mathcal{A}$ and $\mu(conv(S)) \geq \mu(S)$. This proves that $\alpha_3 = \limsup_{S \in \mathcal{A}} \mu(S)$, and hence we establish Theorem \ref{main-theorem}. Finally, Theorem \ref{finite-k} is proved in Section $4$ using the Blaschke selection theorem for the sphere.

\section{Near-optimality of union of a finite number of almost disjoint dyadic cells}

In this section, we prove that $\alpha_3 = \limsup_{S \in \mathcal{B}} \mu(S)$ (Theorem \ref{near-optimal-dyadic-cells}). To be more specific, we show that for 
every orthogonal-pair-free set $S$ and every real number $\epsilon > 0$, there exists an orthogonal-pair-free set $S' \in \mathcal{B}$ such that $\mu(S') > \mu(S) - \epsilon$.

This section is organized as follows. In Section \ref{prelim-measure},
we describe the measure-theoretic background necessary for proving the
main result. In Section \ref{dyadic-decompose}, we describe the dyadic
decompositions $D_k$ of the sphere, where $k$ is a non-negative integer. These decompositions produce a countably infinite set of dyadic cells. In Section \ref{union-of-dense-cells}, we prove Theorem \ref{theorem2}. The theorem shows that for every $\epsilon > 0$ and for any measurable orthogonal-pair-free set $S$, one can construct a finite set $W$ of almost disjoint dyadic cells such that (i) at least $1 - \epsilon$ fraction of the measure of $S$ is concentrated in these cells, and further (ii) for each dyadic cell $c \in W$, $\mu(c \cap S) > (1-\epsilon) \cdot \mu(c)$. 

Note that within each dyadic cell $c \in W$, points of $S$ may be distributed in an arbitrary manner (for example, $S$ may be distributed as a 
fractal, such as the Wallis sieve, inside $c$). We now come to the key idea - the {\it scaling operation}. The scaling operation replaces all the points of $\mu(c \cap S)$, by a completely filled scaled 
dyadic cell $scaled(c) \subset c$. The cell $scaled(c)$ is obtained from $c$ by shrinking its boundary in all directions by an appropriate distance based on the value of $\epsilon$. The scaling operation is described in Section \ref{scaling-operation}. The concluding results in this section are Lemmas \ref{lower-bound-scaled-measure} and \ref{scaled-orthogonal-pair-free}. The first lemma show that  the scaling operation does not reduce the Lebesgue measure by more than a multiplicative factor of $1 - \epsilon$. The second lemma shows that the union of scaled cells is orthogonal-pair-free. Thus, these two lemmas together establish that $\alpha_3 = \limsup_{S \in \mathcal{B}} \mu(S)$.

Finally, we note that while Theorem \ref{union-of-dense-cells} is purely measure-theoretic, the correctness of the scaling operation critically uses the fact that $S$ is orthogonal-pair-free.

\subsection{Preliminaries}
\label{prelim-measure}
\hspace{5mm} {\it Notation.} We use $\sqcup$ to denote the union of disjoint sets.

{\it Anti-podal points.} Two points $p,q \in \mathbb{S}^2$ are called {\it antipodal} if and only if $q=-p$.

{\it Geodesic distance.} For two distinct points $p,q \in \mathbb{S}^2$, such that $p,q$ do not form an antipodal pair, there exists a unique great
circle $G(p,q)$ passing through $p$ and $q$. Points $p$ and $q$ divide $G(p,q)$ into two arcs, of unequal lengths. 
The geodesic distance $d(p,q)$, between points $p$ and $q$, is the length of the smaller arc. Clearly, $0 < d(p,q) < \pi$.

If the two point are antipodal (i.e., $q=-p$), there exist an infinite number of great circles passing
through both $p$ and $q$. Each great circle is divided into two arcs of equal length $\pi$, by these two
points. In this case, we define $d(p,q)=\pi$.
 
Finally, we define $d(p,p)=0$. The geodesic distance $d(p,q)$ forms a metric on the sphere 
$\mathbb{S}^2$. Intuitively, $d(p,q)$ is the minimum distance an ant on the sphere has to travel for
going from point $p$ to point $q$.
 
{\it Geodesic discs.} Let $p$ be a point of the sphere $\mathbb{S}^2$, and let $r$ be a positive real number 
in $(0,\frac{\pi}{2}]$. A geodesic disc on the sphere can be  either open or closed. $B(p,r)$ denotes the open geodesic disc 
of radius $r$ with center at point $p$:

    $$B(p,r) = \{ q ~| ~q \in S^2 ~and ~d(p,q)<r \}$$

$\overline{B(p,r)}$ denotes the closed geodesic disc of radius $r$ with center at point $p$:

          $$\overline{B(p,r)} = \{ q ~| ~q \in \mathbb{S}^2 ~and ~d(p,q) \leq r \}$$

Geodesic discs play the same role in defining measurable sets on the sphere as intervals on the real line, squares in the Euclidean plane, etc.
The area (or, measure) of the geodesic disc is:

$$\mu(B(p,r)) = 2 \pi (1 - \cos(r))$$

{\it Measure theory.} An {\it elementary set} is a finite union of geodesic discs. An {\it open set} $O = \{D_1, D_2, \ldots\}$ 
is a countable union of open geodesic discs $D_1, D_2, \ldots$. Let $M$ be a subset of $\mathbb{S}^2$. 
For a set $M \subset \mathbb{S}^2$, the {\it Lebesgue outer measure} $\mu^*(M)$ is defined as:

    $$\mu^*(M) = \inf_{ O ~| ~O=\{D_1, D_2, \ldots\} ~is ~an ~open ~set ~and ~M \subseteq O} \bigg( \sum_{i \in \mathbb{N}} \mu(D_i) \bigg)$$

In the following, we use one of several equivalent definitions of a measurable set (see \cite{royden}, Chapter $3$, Proposition $15$):

A subset $M$ of $\mathbb{S}^2$ is {\it measurable} if and only if for every $\epsilon>0$, there exists an open set $O$ such that (i) $M$ is a subset of $O$, and (ii) $\mu^*(O-M) < \epsilon$.

We now describe a parametrization of the sphere $\mathbb{S}^2$. Let $H$ be the upper closed hemisphere i.e., the closed geodesic disc of radius
$\frac{\pi}{2}$ with center at the north pole $u=(0,0,1)$. The following map $\psi$ is a bijection from $H$ to the half-open rectangle $R= [0, \frac{\pi}{2}] \times [0, 2 \pi)$ in $\mathbb{R}^2$: $\psi(p) = (\theta, \phi)$, where $(1, \theta, \phi)$ are the spherical polar coordinates of point $p$. 

Let $M$, $M \subseteq H$, be a measurable set on the sphere. We have the following equation:

$$\mu(M) = \int_{(\theta, \phi) \in \psi(M)} \sin(\theta) d \theta d \phi$$

, where the integral is taken with respect to the Lebesgue measure in the Euclidean plane $\mathbb{R}^2$. 

Note that the integral is well-defined because $\sin(\theta)$ is a nonnegative measurable function on rectangle $R$ (see
\cite{royden}, Chapter $3$, Section $5$).

{\it Properties of measurable sets.} If $A$ and $B$ are measurable sets, $A-B$, $A \cup B$, and $A \cap B$ are also measurable sets.  Further, $\mu(A \cup B) \leq \mu(A) + \mu(B)$. If $\mu(A \cap B) = 0$, $\mu(A \cup B) = \mu(A) + \mu(B)$.
If $A \subseteq B$, $\mu(A) \leq \mu(B)$ and $\mu(B-A) = \mu(B) - \mu(A)$. 

Let $A_1, A_2, \ldots$ be a countable family of measurable sets. Then, $A = \cup_{i=1}^{\infty} A_i$ is a measurable set and $\mu(A) \leq \sum_{i=1}^{\infty} \mu(A_i)$.
If $\mu(A_i \cap A_j) = 0$ for all $i \neq j$, $\mu(A) = \sum_{i=1}^{\infty} \mu(A_i)$.

\subsection{Dyadic decomposition}
\label{dyadic-decompose}

For a point $w \in \mathbb{S}^2$, let $G(w)$ denote the great circle polar to 
point $w$ i.e.,

$$G(w) = \bigg\{ w’ ~| ~w’ \in \mathbb{S}^2 ~and ~d(w,w’)=\frac{\pi}{2} \bigg\}$$

Let $m=4 \cdot 4^k$, where $k$ is a non-negative integer.
Let $C_1$ be the circle $\{ (x,y,z) ~| ~x=0 ~and ~y^2+z^2=1 \}$. 
Let $C_2$ be the circle $\{ (x,y,z) ~| ~z=0 ~and ~x^2+y^2=1 \}$.
Note that both $C_1$ and $C_2$ are subsets of sphere $\mathbb{S}^2$.

The dyadic decomposition $D_0$ consists of the $4$ cells of the sphere
$\mathbb{S}^2$ formed by the two circles $C_1$ and $C_2$. $C_1$ is a longitude
and $C_2$ is a latitude, and each cell of $D_0$ is a half-hemisphere. 

For $k \geq 1$, the dyadic decomposition $D_k$ of level $k$ is obtained from the previous
dyadic decomposition $D_{k-1}$ (of level $k-1$) as follows:

\begin{enumerate}
    \item For $k \geq 2$, between any two consecutive longitudes $l_1, l_2$ of $D_{k-1}$, we add a longitude
    bisecting the lune formed by $l_1$ and $l_2$. For $k=1$, we add the circle 
    $C_3=\{ (x,y,z) ~| ~y=0 ~and ~x^2+z^2=1 \}$ to the set of longitudes.
    
    \item Between any two consecutive latitudes $m_1, m_2$ of $D_{k-1}$, we add a latitude
    $m_3$ such that the area enclosed between $m_1$ and $m_3$ is equal to the area enclosed
    between $m_3$ and $m_2$. (We take the north and south poles as the first and last latitudes 
    respectively.)
\end{enumerate}

These latitudes and longitudes of $D_{k}$ together partition the sphere $\mathbb{S}^2$ into exactly 
$4 \cdot 4^k$ cells. Each cell is either a spherical triangle or a spherical quadrilateral.
Note that the longitudinal edges of these triangles and quadrilaterals are geodesic segments, whereas the latitudinal edges are circular arcs on the sphere (unless the latitude is the equator, in which case it is a geodesic segment). 

Note that, for each $k \geq 1$, $D_{k}$ is a refinement of $D_{k-1}$. In fact, each cell of $D_{k-1}$ is partitioned into exactly $4$ cells of equal area in $D_{k}$, and hence all cells of $D_k$ have the same area. To be specific, each cell of $D_k$ has area equal to $\frac{\mu(\mathbb{S}^2)}{4 \cdot 4^k}$. 

We now describe the dyadic decompositions in terms of the parametrization $\psi$ defined above. $\psi(D_k \cap H)$ partitions rectangle $R$ into
a grid $G_k$ formed by $2^{k+1}$ horizontal lines and $2^{k+1}$ vertical lines. The dyadic cells of $D_k$ correspond to rectangular cells of grid $G_k$.
Finally, note that the horizontal lines of $G_k$ are equally spaced, whereas the spacing between vertical lines is variable.

The set $W$ of {\it dyadic cells} consists of all cells $c$ such that $c$ is a cell of dyadic decomposition $D_k$ for some $k \geq 0$. Note that set $W$ is countably infinite. The level of a dyadic cell is the same as the level of its dyadic decomposition. Two dyadic cells $c_1, c_2 \in W$ are {\it almost disjoint} if and only if they do not have a common interior point. 

An alternative definition of open set is as follows  (the spherical counterpart of \cite{tao}, Theorem $1.2.11$): a set $O \subset \mathbb{S}^2$ is 
an open set if and only if $O$ can be written as the countable union of almost disjoint dyadic cells.

The above alternative definition leads to the following lemma:

\begin{lemma}
\label{lemma1}
Let $M$ be a measurable subset of $\mathbb{S}^2$. Then, for every $\epsilon > 0$, there exists an 
integer $k_{\epsilon} \geq 0$ and a subset $W_{\epsilon}$ of dyadic cells of $D_{k_{\epsilon}}$ such that:

\begin{enumerate}
    \item $\mu( M \cap (\cup_{c \in W_{\epsilon}} c) ) > \mu(M) - \epsilon$, and
    \item $\mu( \cup_{c \in W_{\epsilon}} c ) < \mu(M) + \epsilon$. 
\end{enumerate}
\end{lemma} 

\noindent {\bf Proof:} Define $\epsilon_1 = \epsilon_2 = \frac{\epsilon}{2}$. By the definition of a measurable set, there exists an open set $O$ such that (i) $O$ contains $M$, and (ii) $\mu^*(O-M) < \epsilon_1$. Further, since every open set is measurable, $O-M$ is a measurable set, and hence $\mu(O-M)=\mu^*(O-M)$. Suppose $O$ is the countable union of almost disjoint dyadic cells $c_1, c_2, \ldots$, and suppose the 
above cells are ordered in nondecreasing order of their levels.

Since the $c_i$’s are almost disjoint and measurable (i.e., $\mu(c_i \cap c_j) = 0$ for all $i \neq j$), we conclude that:

    $$\mu(O) = \mu(c_1) + \mu(c_2) + \cdots$$ 

Since the right hand side is a convergent infinite sum of positive terms, there exists an index $j^*$ such that:

    $$\mu(c_1) + \mu(c_2) + \cdots + \mu(c_{j*}) > \mu(O) - \epsilon_2$$
    
Let $k^*$ be the maximum dyadic level of any cell in the set $\{c_1, c_2, …, c_{j^*}\}$. 
In fact, due to our ordering above, $k^*$ is the equal to the dyadic level of the last cell $c_{j^*}$.
Define $k_{\epsilon} = k^*$. Further, define $W_{\epsilon}$ as the set 

$$W_{\epsilon} = \{ c ~| ~c ~is ~a ~dyadic ~cell ~of ~D_{k^*} ~and ~c \subseteq c_i ~for ~some ~1 \leq i \leq j^*\}$$

Define $O’ = \cup_{c \in W_{\epsilon}} c$. Note that :
    
    $$\mu(O') = \mu( \cup_{c \in W_{\epsilon}}  c  ) = \mu( \cup_{i=1}^{j^*} c_i ) = \mu(c_1) + \mu(c_2) + \cdots + \mu(c_{j^*})$$

Thus, we conclude that:

     $$\mu(O’) > \mu(O) - \epsilon_2$$

Further, note that ($\sqcup$ denotes disjoint union)

$$\mu(O') = \mu( (O' \cap M) \sqcup (O'- M) ) = \mu(O' \cap M) + \mu(O'-M)$$

Since $O' \subset O$, $O'-M \subset O-M$, and hence $\mu(O'-M) \leq \mu(O-M) < \epsilon_1$.

Thus,

$$\mu(O') < \mu(O' \cap M) + \epsilon_1$$

We conclude that $\mu(O' \cap M) > \mu(O') - \epsilon_1 > \mu(O) - \epsilon_1 - \epsilon_2 = \mu(O) - \epsilon$.

For the upper bound, note that $\mu(O’) \leq \mu(O) = \mu( M \sqcup (O-M)) = \mu(M) + \mu(O-M) < \mu(M) + \epsilon_1 < \mu(M) + \epsilon$.

Hence, the lemma is proved $\blacksquare$

\subsection{Near-optimality of finite union of cells with density greater than $ 1 - \epsilon$}
\label{union-of-dense-cells}

We now prove our next theorem:

\begin{theorem}
\label{theorem2}
Let $M$ be a measurable subset of the sphere such that $\mu(M)>0$. 
Let $\beta=\frac{1}{64}$. Let $0 < \epsilon < \beta$ be a real number.  Then there exists an integer $k_{\epsilon} \geq 0$ 
and a subset $W_{\epsilon}$ of dyadic cells of $D_{k_{\epsilon}}$ such that:

\begin{enumerate}
    \item $\mu( M \cap (\cup_{c \in W_{\epsilon}} c) ) > (1 - \epsilon) \cdot \mu(M)$, and
    \item for each cell $c \in W_{\epsilon}$:
                $$\mu(c \cap M) \geq (1-\epsilon) \cdot \mu(c)$$
\end{enumerate}
\end{theorem}

\noindent {\bf Proof:} Let $\epsilon_1  = \frac{\epsilon^3}{27}$. Define $\epsilon_2 = \epsilon_1 \cdot \mu(M)$.
By Lemma \ref{lemma1}, there exists an integer $k_{\epsilon_2} \geq 0$ and a subset $W_{\epsilon_2}$ of dyadic cells of $D_{k_{\epsilon_2}}$ such that:

$$\mu( M \cap (\cup_{c \in W_{\epsilon_2}} c) ) > \mu(M) - \epsilon_2$$
$$= (1 - \epsilon_1) \cdot \mu(M)$$ 

and,

$$\mu( \cup_{c \in W_{\epsilon_2}} c ) < \mu(M) + \epsilon_2$$
$$= (1 + \epsilon_1) \cdot \mu(M)$$

Let $W’$ be the subset of all cells $c$ of $W_{\epsilon_2}$ such that:

 $$\mu(c \cap M) \geq (1 - \sqrt[3]{\epsilon_1} ) \cdot \mu(c)$$

Note that, for all cells $c \in W'$, $\mu(c \cap M) \geq \bigg(1 - \sqrt[3]{\frac{\epsilon^3}{27}} \bigg) \cdot \mu(c) = ( 1 - \frac{\epsilon}{3} ) \cdot \mu(c) > ( 1 - \epsilon) \cdot \mu(c)$.

Let $n$ be the number of cells in $W_{\epsilon_2}$. We repeatedly use the fact that all dyadic cells at the same level are of equal area. Therefore, if $W_{\epsilon_2}$ has $n$ dyadic cells, then for any cell
$c \in W_{\epsilon_2}$:

    $$\mu(\cup_{c \in W_{\epsilon_2}} c) = n \cdot \mu(c)$$

\begin{observation}
\label{obs-A1}
        $$1 - \epsilon_1^{\frac{5}{6}} + \epsilon_1 < 1 - \epsilon_1$$
\end{observation}

\noindent {\bf Proof:} Let $r > 0$ be the unique real number such that $\epsilon_1 = \frac{1}{r^{6}}$. Note that $\epsilon_1 = \frac{\epsilon^3}{27} < \epsilon < \beta = \frac{1}{2^6}$. Thus, $r > 2$. Then,
        
        $$1 - \epsilon_1^{\frac{5}{6}} + \epsilon_1$$
        $$= 1 - \frac{1}{r^{5}} + \frac{1}{r^6}$$
        $$= 1 - \bigg( \frac{1}{r^6} \cdot (r-1) \bigg)$$
        $$ < 1 - \frac{1}{r^6} = 1 - \epsilon_1$$

$\blacksquare$

\begin{observation}
At least $(1 - \sqrt{\epsilon_1}) \cdot n$ cells of $W_{\epsilon_2}$ belong 
to $W’$. In other words, $\mu(W') \geq (1 - \sqrt{\epsilon_1}) \cdot n \cdot \mu(c)$.
\end{observation}

\noindent {\bf Proof:} Suppose, for the sake of contradiction, that strictly less than $(1 - \sqrt{\epsilon_1}) \cdot n$ cells of $W_{\epsilon_2}$ belong 
to $W’$. Then,

    $$\mu( M \cap (\cup_{c \in W_{\epsilon_2}} c) )$$
    $$\leq \mu ( M \cap (\cup_{c \in W’} c) ) + \mu( M \cap (\cup_{c \in W_{\epsilon_2}-W’} c) ) ~(since ~W_{\epsilon_2} = W' \sqcup (W_{\epsilon_2} - W')~)$$
    
    $$ = \mu ( \cup_{c \in W'} (c \cap M) ) + \mu( \cup_{c \in W_{\epsilon_2}-W’} (c \cap M) )$$

    $$= \sum_{c \in W'} \mu(c \cap M) + \sum_{c \in W_{\epsilon_2}-W’} \mu(c \cap M)$$

    $$\leq \sum_{c \in W'} \mu(c) + \sum_{c \in W_{\epsilon_2}-W’} \mu(c \cap M) ~(since ~\mu(c \cap M) \leq \mu(c) ~)$$

    $$ < \mu(W') + \sum_{c \in W_{\epsilon_2}-W’} ((1 - \sqrt[3]{\epsilon_1} ) \cdot \mu(c)) ~(since ~\mu(c \cap M) < (1 - \sqrt[3]{\epsilon_1} ) \cdot \mu(c) ~for ~c \in W_{\epsilon_2} - W' ~)$$
    
    $$ = \mu(W') + (1 - \sqrt[3]{\epsilon_1} ) \cdot \mu(W_{\epsilon_2}-W’)$$

    $$=  \mu(W') + (1 - \sqrt[3]{\epsilon_1} ) \cdot (\mu(W_{\epsilon_2})-\mu(W’)) ~(since ~W' \subset W_{\epsilon_2} ~)$$

    $$= \sqrt[3]{\epsilon_1} \cdot \mu(W') + (1 - \sqrt[3]{\epsilon_1} ) \cdot \mu(W_{\epsilon_2})$$

    $$< \sqrt[3]{\epsilon_1} \cdot n \cdot (1-\sqrt{\epsilon_1}) \cdot \mu(c) + (1 - \sqrt[3]{\epsilon_1} ) \cdot n \cdot \mu(c)$$
    
    $$= n \cdot \mu(c) - n \cdot \epsilon_1^{\frac{5}{6}} \cdot \mu(c)$$
    $$= (1-\epsilon_1^{\frac{5}{6}}) \cdot n \cdot \mu(c)$$
    $$= (1-\epsilon_1^{\frac{5}{6}}) \cdot \mu( \cup_{c \in W_{\epsilon_2}} c)$$
    $$< (1-\epsilon_1^{\frac{5}{6}}) \cdot (1 + \epsilon_1) \cdot \mu(M)$$
    $$= ( 1 - \epsilon_1^{\frac{5}{6}} + \epsilon_1 - \epsilon_1^{\frac{11}{6}} ) \cdot \mu(M)$$
    $$< ( 1 - \epsilon_1^{\frac{5}{6}} + \epsilon_1 ) \cdot \mu(M)$$
    
By Observation \ref{obs-A1}:

    $$\mu( M \cap (\cup_{c \in W_{\epsilon_2}} c) ) < ( 1 - \epsilon_1) \cdot \mu(M)$$

We arrive at a contradiction, and hence the observation is proved $\blacksquare$

Thus,

$$\mu( M \cap (\cup_{c \in W’} c) ) = \mu( \cup_{c \in W'} (c \cap M) ) = \sum_{c \in W'} \mu(c \cap M)$$    

$$ \geq \sum_{c \in W'} ((1 - \sqrt[3]{\epsilon_1} ) \cdot \mu(c))$$
$$= (1 - \sqrt[3]{\epsilon_1} ) \cdot \mu(W')$$
$$\geq (1 - \sqrt[3]{\epsilon_1} ) \cdot (1 - \sqrt{\epsilon_1}) \cdot n \cdot \mu(c)$$
$$= (1 - \sqrt{\epsilon_1} - \sqrt[3]{\epsilon_1} + \epsilon_1^{\frac{5}{6}}) \cdot n \cdot \mu(c)$$
$$> (1 - \sqrt{\epsilon_1} - \sqrt[3]{\epsilon_1}) \cdot n \cdot \mu(c)$$
$$> (1 - 2 \cdot \sqrt[3]{\epsilon_1}) \cdot n \cdot \mu(c)  ~(since ~\epsilon_1 < \beta < 1, ~\sqrt{\epsilon_1} < \sqrt[3]{\epsilon_1})$$

$$= (1 - 2 \cdot \sqrt[3]{\epsilon_1}) \cdot \mu( \cup_{c \in W_{\epsilon_2}} c)$$
$$\geq  (1 - 2 \cdot \sqrt[3]{\epsilon_1}) \cdot \mu( M \cap (\cup_{c \in W_{\epsilon_2}} c) ) $$
 $$   >  (1 - 2 \cdot \sqrt[3]{\epsilon_1}) \cdot (1-\epsilon_1) \cdot \mu(M)$$
$$   = (1 - 2 \cdot \sqrt[3]{\epsilon_1} - \epsilon_1 + 2 \cdot \epsilon_1^{\frac{4}{3}}) \cdot \mu(M)$$
 $$   > (1 - 2 \cdot \sqrt[3]{\epsilon_1} - \epsilon_1) \cdot \mu(M) $$
$$    > (1 - 3 \cdot \sqrt[3]{\epsilon_1}) \cdot \mu(M) ~(since ~\epsilon_1 < \beta < 1, ~\epsilon_1 < \sqrt[3]{\epsilon_1})$$

$$ = (1 - \epsilon) \cdot \mu(M) ~\bigg(since ~\sqrt[3]{\epsilon_1} = \frac{\epsilon}{3}~\bigg)$$

Take

    $$W_{\epsilon}=W’$$

to derive the theorem $\blacksquare$

\subsection{Near-optimality of finite union of scaled dyadic cells}
\label{scaling-operation}
Let $M$ be any measurable, orthogonal-pair-free set. Let $0 < \epsilon < 1$ be a real number. Define $\delta = \sqrt[2]{\frac{\epsilon \cdot \mu(M)}{4 \pi}}$.
Let $D^+$ and $D^-$ be two geodesic discs of radius $\delta$, with centers 
at the north and south pole respectively. Before proceeding with the scaling operation, we remove the portion of set $M$ which
lies in $D^+ \cup D^-$. The removal of $M \cap (D^+ \cup D^-)$ allows us to prove
Observation \ref{obs-lb-area-scaled-cell}.

Define a new set $M_1 = M - (D^- \cup D^+)$. Since $M_1 \subseteq M$, $M_1$ is also orthogonal-pair-free.
Further, $\mu(M_1) = \mu(M) - \mu(D^- \cup D^+) = \mu(M) - 2 \cdot 2 \pi \cdot (1 - \cos(\delta))$. Since
$1 - \cos(\delta) \leq \frac{\delta^2}{2}$ for $\delta \in \big[0, \frac{\pi}{2}\big]$, we conclude that $\mu(M_1) \geq \mu(M) - 4 \pi \cdot \frac{\delta^2}{2} = \mu(M) - 2 \pi \delta^2 = \mu(M)-\frac{\epsilon}{2} \cdot \mu(M) = \big( 1 - \frac{\epsilon}{2} \big) \cdot \mu(M)$.

Before going further, we list some trigonometric facts and describe the choice of constants.

\subsubsection{Preliminaries}
\label{preliminaries-trig}

{\it Plane trigonometry.} For $0 \leq x \leq \frac{\pi}{2}$, $\frac{x}{2} \leq \sin(x) \leq x$ and $1 - \cos(x) \leq \frac{x^2}{2}$. For $0 \leq x \leq 1$, $1-\cos(x) \geq \frac{x^2}{2} - \frac{x^4}{4!}$.

\noindent {\it Spherical trigonometry.} Let $L$ be a lune formed by two great circles $g_1$ and $g_2$. Let $\theta$ be the angle of the lune. Let $0 < r < \frac{\pi}{2}$. Let $p$ be a point on great circle $g_1$ at distance $r$ from a vertex $v$ of $L$. Let $q \in g_2$ be a point such that $d(p,q) = d(p,g_2)$. Consider the spherical triangle $\triangle pqv$. Note that the angle at vertex $q$ of the triangle is a right angle.
By the spherical law of sines, $\frac{\sin(d(p,g_2))}{\sin(\theta)} = \frac{\sin(r)}{\sin \big(\frac{\pi}{2} \big)}$. We conclude that $\sin(\theta) = \frac{\sin(d(p,g_2))}{\sin(r)}$.

\subsubsection{Choice of constants} 
\label{choice-of-constants}

Let $N=3$ and let
$\epsilon_1 = \epsilon^6$. Choose $\epsilon > 0$ such that the following two equations are satisfied:

$$\bigg( 1 - \epsilon_1^{\frac{1}{N}} \cdot \bigg( 3 \sqrt[2]{\pi} + \frac{2}{\sin(\delta) \cdot \sqrt[2]{\pi}} \bigg) \bigg) \cdot \bigg( 1 - \epsilon_1 \bigg) \cdot \bigg( 1 - \frac{\epsilon}{2} \bigg) \geq 1 - \epsilon$$

and,

$$\sin \bigg( \frac{\pi}{8} \bigg) \cdot \bigg(1 - \frac{16 \cdot \sqrt{2}}{\sqrt{\pi}} \cdot \epsilon_1^{1 - \frac{2}{N}} \bigg) \cdot \epsilon_1^{\frac{2}{N}} > 2 \cdot \epsilon_1$$

{\it Note.} Such a choice of $\epsilon_1 > 0$ is possible since

$$\lim_{\epsilon \rightarrow 0} \frac{ \epsilon^2 \cdot \Bigg( 3 \sqrt[2]{\pi} + \frac{2}{\sin \big(\sqrt[2]{\frac{\epsilon \cdot \mu(M)}{4 \pi}} \big) \cdot \sqrt[2]{\pi}} \Bigg)}{\epsilon} = 0$$

Hence, for sufficiently small $\epsilon > 0$ (equivalently, sufficiently small $\epsilon_1 > 0$, since $\epsilon = \sqrt[6]{\epsilon_1}$):

$$1 - \epsilon_1^{\frac{1}{N}} \cdot \bigg( 3 \sqrt[2]{\pi} + \frac{2}{\sin(\delta) \cdot \sqrt[2]{\pi}} \bigg) \geq 1 - \frac{\epsilon}{4}$$

Thus, for sufficiently small $\epsilon > 0$, 

$$\bigg( 1 - \epsilon_1^{\frac{1}{N}} \cdot \bigg( 3 \sqrt[2]{\pi} + \frac{2}{\sin(\delta) \cdot \sqrt[2]{\pi}} \bigg) \bigg) \cdot \bigg( 1 - \epsilon_1 \bigg) \cdot \bigg( 1 - \frac{\epsilon}{2} \bigg)$$

$$\geq \bigg( 1 - \frac{\epsilon}{4} \bigg) \cdot \bigg( 1 - \epsilon^6 \bigg) \cdot \bigg( 1 - \frac{\epsilon}{2} \bigg) \geq 1 - \epsilon$$

Further,

$$\lim_{\epsilon_1 \rightarrow 0} \frac{\sin \big( \frac{\pi}{8} \big) \cdot \bigg(1 - \frac{16 \cdot \sqrt{2}}{\sqrt{\pi}} \cdot \epsilon_1^{\frac{1}{3}} \bigg) \cdot \epsilon_1^{\frac{2}{3}} }{2 \cdot \epsilon_1} = \lim_{\epsilon_1 \rightarrow 0} \frac{\sin \big( \frac{\pi}{8} \big) \cdot \epsilon_1^{\frac{2}{3}}}{2 \cdot \epsilon_1} - \frac{ \sin \big( \frac{\pi}{8} \big) \cdot 8 \cdot \sqrt{2}}{\sqrt{\pi}} = \infty$$

Hence, for sufficiently small $\epsilon > 0$ (equivalently, sufficiently small $\epsilon_1 > 0$, since $\epsilon = \sqrt[6]{\epsilon_1}$):

$$\sin \bigg( \frac{\pi}{8} \bigg) \cdot \bigg(1 - \frac{16 \cdot \sqrt{2}}{\sqrt{\pi}} \cdot \epsilon_1^{1 - \frac{2}{N}} \bigg) \cdot \epsilon_1^{\frac{2}{N}} > 2 \cdot \epsilon_1$$

\subsubsection{The scaling operation}
 
We now construct a new set $M_2$ as follows. Apply Theorem \ref{theorem2} on set $M_1$, using the value of $\epsilon_1$ from Section \ref{choice-of-constants}. 

Let $W_{\epsilon_1}$ be the set of dyadic cells of $D_{k_{\epsilon_1}}$ generated as a result.
For a cell $c$, let $bd(c)$ denote the boundary of $c$. For a point $z$ on the sphere, let $d(z,bd(c))$ 
denote the distance of $z$ from $bd(c)$ i.e.,

        $$d(z,bd(c)) = \inf_{z’ \in bd(c)} d(z,z’)$$

Recall from Section \ref{choice-of-constants} that $N=3$. For each cell 
$c \in W_{\epsilon_1}$, construct a closed region $scaled(c)$ as follows:

$$scaled(c) = \{ y ~| ~y \in c ~and ~d(y, bd(c)) \geq \epsilon_1^{\frac{1}{N}} \cdot \sqrt{\mu(c)} \}$$

Define $M_2$ to be the union of regions $scaled(c)$, where $c$ varies over dyadic cells in
$W_{\epsilon_1}$. Note that $M_1$ is not necessarily a subset of $M$. Further, let $A = \cup_{c \in W_{\epsilon_1}} c$.

\subsubsection{Lebesgue measure of union of scaled dyadic cells}

\begin{observation}
\label{obs-lb-area-scaled-cell}
For each cell $c$ in $W_{\epsilon_1}$, $scaled(c)$ is measurable and

        $$\mu(scaled(c)) \geq \bigg( 1 - \epsilon_1^{\frac{1}{N}} \cdot \bigg( 3 \sqrt[2]{\pi} + \frac{1}{\sin(\delta) \cdot \sqrt[2]{\pi}} \bigg) \bigg) \cdot \mu(c)$$

Further, for any two points $x,y \in scaled(c)$, there exists a finite piecewise-linear 
curve $T(x,y)$ connecting $x$ and $y$, such that $T(x,y) \subset scaled(c)$.
\end{observation}

\noindent {\bf Proof:}  Suppose dyadic cell $c$ is at level $k$. Then, $\mu(c) = \frac{4 \pi}{4 \cdot 4^k}$. Further, the angle $\omega$ between the two great circles containing the two vertical sides of $c$ is $\frac{2 \pi}{2 \cdot 2^k}$.
Let $C_1$ and $C_2$ be the circles containing the two horizontal sides of $c$. 
Let $r_1, r_2$ ($r_1 < r_2$) be the radii of $C_1$ and $C_2$, respectively.

Since $c$ does not intersect $D^+ \cup D^-$, the $\delta \leq r_1, r_2 \leq \frac{\pi}{2}$. Thus, lengths of the two horizontal circular arcs lie in the interval $[\omega \cdot \sin(\delta), \omega \cdot \sin \big( \frac{\pi}{2} \big)]$.

The area of the spherical strip $S$ enclosed by $C_1$ and $C_2$ is $\frac{4 \pi}{2 \cdot 2^k}$.  Further, $\mu(S) = 2 \pi \cdot ( \cos(r_1) - \cos(r_2) )$. Hence,
$\cos(r_1) - \cos(r_2) = \frac{\mu(S)}{2 \pi} = \frac{2 \cdot 2^k \cdot \mu(c)}{2 \pi} = \frac{\mu(c)}{\omega}$.

Let $\epsilon_2 = \epsilon_1^{\frac{1}{N}} \cdot \sqrt{\mu(c)}$.

If we increase the radius of $D_1$ by $\epsilon_2$, the measure of region $B_1$ removed from cell $c$ is $\omega \cdot ( \cos(r_1) - \cos(r_1 + \epsilon_2))$. Similarly, if we decrease the radius of $D_2$ by $\epsilon_2$, the measure of region $B_2$ removed from cell $c$ is $\omega \cdot ( \cos(r_2-\epsilon_2) - \cos(r_2))$.

Let $\omega'$ be a real number such that $\sin(\omega') = \frac{\sin(\epsilon_2)}{\sin(r)}$. Suppose we decrease the angle $\omega$ at north pole by $\omega'$ equally on both sides. The measure of region $B_3$ removed from cell $c$ by the inward rotation of left vertical edge is $\omega' \cdot (\cos(r_1) - \cos(r_2))$. Further, the measure of
region $B_4$ removed from cell $c$ by the inward
rotation of right vertical edge is the same as $\mu(B_3)$.

 Let $e_1, e_2, e_3, e_4$ be the top, bottom, left, and right sides of $c$. Observe that $d(c-B_1, e_1) \geq \epsilon_2$, $d(c-B_2, e_2) \geq \epsilon_2$, $d(c-B_3, e_3) \geq \epsilon_2$, and $d(c-B_4, e_4) \geq \epsilon_2$, where the last two inequalities are due to the property of spherical lunes noted in Section \ref{preliminaries-trig}. Thus, we conclude that $c - (B_1 \cup B_2 \cup B_3 \cup B_4) \subseteq scaled(c)$.

Therefore, $\mu(scaled(c)) \geq \mu(c) - \mu(B_1 \cup B_2 \cup B_3 \cup B_4)
\geq \mu(c) - \mu(B_1) - \mu(B_2) - \mu(B_3) - \mu(B_4)$ (by finite subadditivity of measure). 

Note that $\mu(B_1) \leq \omega \cdot ( (1 - \cos(\epsilon_2)) \cdot \cos(r_1) + \sin(\epsilon_2) \cdot \sin(r_1)) \leq \omega \cdot ( \frac{\epsilon_2^2}{2} \cdot \cos(r_1) + \epsilon_2 \cdot \sin(r_1)) \leq \omega \cdot ( \frac{\epsilon_2^2}{2} + \epsilon_2) \leq \omega \cdot 2 \epsilon_2$.

Similarly, $\mu(B_2) \leq \omega \cdot ( \sin(r_2) \sin(\epsilon_2) - (1 - \cos(\epsilon_2)) \cdot \cos(r_2)) \leq \omega \cdot \sin(r_2) \sin(\epsilon_2)
\leq \omega \cdot \epsilon_2$.

Further, $\frac{\omega'}{2} \leq \sin(\omega') = \frac{\sin(\epsilon_2)}{\sin(r_1)} \leq \frac{\epsilon_2}{\sin(r_1)} \leq \frac{\epsilon_2}{\sin(\delta)}$. Thus, $\mu(B_3) \leq \frac{2 \cdot \epsilon_2}{\sin(\delta)} \cdot (\cos(r_1) - \cos(r_2)) = 2 \cdot \frac{\epsilon_2}{\sin(\delta)} \cdot \frac{\mu(c)}{\omega}$.

Thus,

$$\mu(scaled(c)) \geq \mu(c) - 3 \cdot \omega \cdot \epsilon_2 - 2 \cdot \frac{\epsilon_2}{\sin(\delta)} \cdot \frac{\mu(c)}{\omega}$$

Since $\omega = \sqrt[2]{\pi} \cdot \sqrt[2]{\mu(c)}$, by the above equation:

$$\mu(scaled(c)) \geq \mu(c) - 3 \cdot \sqrt[2]{\pi} \sqrt[2]{\mu(c)} \cdot \epsilon_1^{\frac{1}{N}} \sqrt[2]{\mu(c)} - 2 \cdot \frac{\epsilon_1^{\frac{1}{N}} \cdot \sqrt[2]{\mu(c)}}{\sin(\delta)} \cdot \frac{\sqrt[2]{\mu(c)}}{\sqrt[2]{\pi}}$$

$$= \mu(c) \cdot \bigg( 1 - \epsilon_1^{\frac{1}{N}} \cdot \bigg( 3 \sqrt[2]{\pi} + \frac{2}{\sin(\delta) \cdot \sqrt[2]{\pi}} \bigg) \bigg)$$
$\blacksquare$

\begin{lemma}
\label{lower-bound-scaled-measure}
$$\mu(M_2) \geq (1 - \epsilon) \cdot \mu(M)$$
\end{lemma}
        
\noindent {\bf Proof:}

        $$\mu(M_2) = \sum_{c \in W_{\epsilon_1}} \mu(scaled(c))$$

        $$ \geq \sum_{c \in W_{\epsilon_1}} \bigg( \bigg( 1 - \epsilon_1^{\frac{1}{N}} \cdot \bigg( 3 \sqrt[2]{\pi} + \frac{2}{\sin(\delta) \cdot \sqrt[2]{\pi}} \bigg) \bigg) \cdot \mu(c) \bigg)$$

        $$= \bigg( 1 - \epsilon_1^{\frac{1}{N}} \cdot \bigg( 3 \sqrt[2]{\pi} + \frac{2}{\sin(\delta) \cdot \sqrt[2]{\pi}} \bigg) \bigg)\cdot \sum_{c \in W_{\epsilon_1}} \mu(c)$$

        $$= \bigg( 1 - \epsilon_1^{\frac{1}{N}} \cdot \bigg( 3 \sqrt[2]{\pi} + \frac{2}{\sin(\delta) \cdot \sqrt[2]{\pi}} \bigg) \bigg)\cdot \mu(A)$$

         $$\geq \bigg( 1 - \epsilon_1^{\frac{1}{N}} \cdot \bigg( 3 \sqrt[2]{\pi} + \frac{2}{\sin(\delta) \cdot \sqrt[2]{\pi}} \bigg) \bigg)\cdot \mu(A \cap M_1)$$

          $$> \bigg( 1 - \epsilon_1^{\frac{1}{N}} \cdot \bigg( 3 \sqrt[2]{\pi} + \frac{2}{\sin(\delta) \cdot \sqrt[2]{\pi}} \bigg) \bigg)\cdot (1 - \epsilon_1) \cdot \mu(M_1)$$
          
          $$\geq \bigg( 1 - \epsilon_1^{\frac{1}{N}} \cdot \bigg( 3 \sqrt[2]{\pi} + \frac{2}{\sin(\delta) \cdot \sqrt[2]{\pi}} \bigg) \bigg)\cdot (1 - \epsilon_1) \cdot \bigg( 1 - \frac{\epsilon}{2} \bigg) \cdot \mu(M)$$

        By choice of constants (Section \ref{choice-of-constants}), this is at least
        $(1 - \epsilon) \cdot \mu(M)$, and hence the claim is proved $\blacksquare$

\subsubsection{Union of scaled dyadic cells is orthogonal-pair-free}

\begin{theorem}
\label{scaled-orthogonal-pair-free}
$M_2$ is orthogonal-pair-free.
\end{theorem}
        
\noindent {\bf Proof:} Suppose, for the sake of contradiction, that 
$M_2$ is not orthogonal-pair-free. Then, there exist two distinct cells 
$c_1, c_2 \in W_{\epsilon_1}$ such that there exists a point $y \in scaled(c_1)$ for which $G(y) \cap scaled(c_2) \neq \phi$.

By the property of our dyadic decomposition, $\mu(c_1) = \mu(c_2)$.
Let $r_1 = \epsilon_1^{\frac{1}{N}} \cdot \sqrt{\mu(c_1)}$.
Let $D_1=B(y, r_1)$ be the open disc of radius $r_1$ with its center at 
point $y$. 
Since $y$ belongs to $scaled(c_1)$, we conclude that $B(y,r_1)$ lies 
inside the cell $c_1$. Note that:

    $$\mu(D_1) = 2 \pi \cdot \bigg( 1 - \cos\bigg(\epsilon_1^{\frac{1}{N}} \cdot \sqrt[2]{\mu(c_1)}\bigg) \bigg)$$

Now, observe that:

    $$\mu(D_1 \cap M) = \mu(D_1) - \mu(D_1 - M)$$
    $$\geq \mu(D_1) - \mu(c_1 - M)$$
    $$\geq \mu(D_1) - \epsilon_1 \cdot \mu(c_1) ~(By ~Theorem ~\ref{theorem2})$$

Thus,

   $$\frac{\mu(D_1 \cap M)}{\mu(D_1)} \geq 1 - \epsilon_1 \cdot \frac{\mu(c_1)}{\mu(D_1)}$$

Note that $\sin(z) \geq \frac{z}{2}$, for $0 \leq z \leq \frac{\pi}{2}$.
Hence, $1 - \cos(y) = 2 \cdot \big( \sin \big( \frac{y}{2} \big) \big)^2 \geq 2 \cdot \big( \frac{y}{4} \big)^2 = \frac{y^2}{8}$. Thus,

Thus,

   $$\mu(D_1) \geq  2 \pi \cdot \frac{\bigg(\epsilon_1^{\frac{1}{N}} \cdot \sqrt[2]{\mu(c_1)}\bigg)^2}{8} = \frac{\pi \cdot \epsilon_1^{\frac{2}{N}}}{4} \cdot \mu(c_1)$$

Further, we can conclude that:

    $$\mu(D_1 \cap M) \geq 1 - \epsilon_1 \cdot \frac{\mu(c_1)}{ \frac{\pi \cdot \epsilon_1^{\frac{2}{N}}}{4} \cdot \mu(c_1)}$$

    $$= 1 - \frac{4}{\pi} \cdot \epsilon_1^{1 - \frac{2}{N}}$$

Parameterize the disk $D_1$ about its center by $r$ 
and $\theta$, where $r$ varies from $-r_1$ to $r_1$, and 
$\theta$ varies from $0$ to $\pi$. We assume that $\theta$ varies along the unit circle $\mathbb{S}^1$, parameterized as $[0,\pi)$.

Let $s(\theta)$ be the geodesic segment $\{ (r,\theta) ~| ~-r_1 \leq r \leq r_1\}$. The set $\cup_{z \in s(\theta)} G(z)$ is a lune $L(\theta)$
of angle $2 \cdot r_1$. The great circle $G(y)$ bisects $L(\theta)$. Let $v_1(\theta)$ and $v_2(\theta)$ be the 
two vertices of lune $L(\theta)$. Clearly, $v_1(\theta)$ and $v_2(\theta)$
are diametrically opposite points on the great circle $G(y)$. As $\theta$ 
moves along the unit circle $[0, \pi)$, the points $v_1(\theta), v_2(\theta)$ rotate along the great circle $G(y)$.

\begin{observation}
\label{obs-dist-from-lune-vertices}
There exists an interval $[\theta_1, \theta_2] \subset [0, \pi)$ such that $\mu([\theta_1, \theta_2]) = \frac{\pi}{2}$ and for each $\theta \in [\theta_1, \theta_2]$:

        $$\min(d(v_1(\theta), scaled(c_2)), d(v_2(\theta), scaled(c_2))) \geq \frac{\pi}{8}$$
\end{observation}

\noindent {\bf Proof:} Choose a sufficiently small $\epsilon > 0$ so that diameter $diam(c_2)$ of cell $c_2$ is less than $\frac{\pi}{8}$. Let $w \in G(y) \cap scaled(c_2) \subset G(y) \cap c_2$. Let $z_1, z_2$ be the two points of $G(y)$ such that $d(w,z_1) = d(w, z_2) = \frac{\pi}{4}$. Let $z'_1, z'_2$ be antipodal points to $z_1, z_2$ respectively. 

Take $[\theta_1, \theta_2]$ as the set of all $\theta$ such that both $v_1(\theta)$ and $v_2(\theta)$ belong to $\overline{z_1 z'_2} \cup \overline{z_2 z'_1}$. Note that, for two points $p,q$ such that $p \neq -q$, $\overline{pq}$ denotes the shortest geodesic segment joining $p$ and $q$. Further, $\overline{z_1 z'_2} \cup \overline{z_1 z'_2} \subset G(y)$.

Clearly, $\mu([\theta_1, \theta_2]) = \frac{\pi}{2}$. Further, by triangle inequality, for any $\theta \in [\theta_1, \theta_2]$, $d(v_1(\theta), scaled(c_2)) \geq d(v_1(\theta), c_2) \geq d(v_1(\theta), w) - diam(c_2) \geq \frac{\pi}{4} - \frac{\pi}{8} = \frac{\pi}{8}$. The same holds for $v_2(\theta)$ and hence the claim is proved
$\blacksquare$

Let $S = \{ (r, \theta) ~| ~\theta \in [\theta_1, \theta_2], ~-r_1 \leq r \leq r_1\}$. Note that

$$\frac{\mu(S \cap M)}{\mu(S)} \geq \frac{\mu(S) - \mu(S-M)}{\mu(S)}$$
$$= 1 - \frac{\mu(S-M)}{\mu(S)}$$
$$\geq 1 - \frac{\mu(D-M)}{\mu(S)} ~(since ~S \subset D, ~S-M \subset D-M)$$

$$\geq 1 - \frac{\mu(D-M)}{\frac{\mu(D)}{8}} ~(since ~\mu(S) \geq \frac{\mu(D)}{8})$$

$$= 1 - 8 \cdot \frac{\mu(D-M)}{\mu(D)}$$

$$ \geq 1 - 8 \cdot \frac{4}{\pi} \cdot \epsilon_1^{1 - \frac{2}{N}}$$

$$ = 1 - \frac{32}{\pi} \cdot \epsilon_1^{1 - \frac{2}{N}}$$

Define $\epsilon_2 = \frac{32}{\pi} \cdot \epsilon_1^{1 - \frac{2}{N}}$.
Let $I_M$ be the indicator function 
of set $M$ i.e., $I_M(z)=1$ if $z \in M$, and is $0$ otherwise.
Note that

$$\mu(S \cap M) = \int_{(r,\theta) \in S} I_M(r, \theta) \cdot |\sin(r)| \cdot da$$

, where $da$ is the Lebesgue measure in $\mathbb{R}^2$.

Define $J_M(r,\theta) = I_M(r, \theta) \cdot |\sin(r)|$. Since $J_M(\cdot)$ is a non-negative measurable function on $S$, by Tonelli's theorem \cite{royden}, we conclude that:

\begin{enumerate}

\item for almost all $\theta \in [\theta_1, \theta_2]$, $f_{\theta}(r)=J_M(r, \theta)$ is a measurable function on $[-r_1, r_1]$,

\item $\int_{-r_1}^{r_1} f_{\theta}(r) dr$ is a measurable function on $[\theta_1, \theta_2]$, and

\item $$\mu(S \cap M) =$$

$$ \int_{\theta=\theta_1}^{\theta_2} \bigg( \int_{r=-r_1}^{r_1} \bigg( I_M(r, \theta) \cdot |\sin(r)| \cdot dr \bigg) \bigg) \cdot d \theta$$
\end{enumerate}

We use the fact that $\frac{\mu(S \cap M)}{\mu(S)} \geq 1 - \epsilon_2$ to conclude that

$$(1 - \epsilon_2) \cdot 2 \cdot (\theta_2 - \theta_1) \cdot (1 - \cos(r_1)) \leq \mu(S \cap M) = \int_{\theta=\theta_1}^{\theta_2} \bigg( \int_{r=-r_1}^{r_1} \bigg( J_M(r, \theta) \cdot dr \bigg) \bigg) \cdot d \theta$$

\begin{observation}
\label{obs-tonelli-appl}
There exists an angle $\theta^* \in [\theta_1, \theta_2]$, such that:

\begin{enumerate}
\item $\int_{r=-r_1}^{r_1} J_M(r, \theta^*) \cdot dr$ exists, and 

\item $\int_{r=-r_1}^{r_1} J_M(r, \theta^*) \cdot dr$ is at least $(1-\epsilon_2) \cdot 2 \cdot (1 - \cos(r_1))$.
\end{enumerate}
\end{observation}

\noindent {\bf Proof:} By Tonelli's theorem \cite{royden}, for almost all $\theta \in [\theta_1, \theta_2]$, $f_{\theta}(r)=J_M(r, \theta)$ is a measurable function on $[-r_1, r_1]$. Let $Z \subset [\theta_1, \theta_2]$ be the set of all $\theta$ such that $f_{\theta}(r)$ is not a measurable function on $[-r_1, r_1]$. Thus, $\mu(Z)=0$. Further,

$$(1 - \epsilon_2) \cdot 2 \cdot (\theta_2 - \theta_1) \cdot (1 - \cos(r_1)) \leq \int_{\theta \in [\theta_1, \theta_2] - Z} \bigg( \int_{r=-r_1}^{r_1} \bigg( J_M(r, \theta) \cdot dr \bigg) \bigg) \cdot d \theta$$

Suppose, for the sake of contradiction, that for all $\theta \in [\theta_1, \theta_2] - Z$, $\int_{r=-r_1}^{r_1} J_M(r, \theta) \cdot dr < (1-\epsilon_2) \cdot 2 \cdot (1 - \cos(r_1))$.

Then, 

$$\int_{\theta \in [\theta_1, \theta_2] - Z} \bigg( \int_{r=-r_1}^{r_1} \bigg( J_M(r, \theta) \cdot dr \bigg) \bigg) \cdot d \theta < \int_{\theta \in [\theta_1, \theta_2] - Z}  (1-\epsilon_2) \cdot 2 \cdot (1 - \cos(r_1)) \cdot d \theta$$

$$= (1 - \epsilon_2) \cdot 2 \cdot (\theta_2 - \theta_1) \cdot (1 - \cos(r_1))$$

We arrive at a contradiction. Thus, there exists a $\theta^* \in [\theta_1, \theta_2] - Z$ for which the above claim is true
$\blacksquare$

Let $t^*$ be the geodesic segment consisting of points 
$\{ z ~| ~z=(r, \theta^*), ~-r_1 \leq r \leq r_1\}$. Note that $t^*$ is a geodesic diameter 
of disc $D_1$. We now prove the following observations:

\begin{observation}
\label{J-lower-bound}
Let $A$ be a measurable subset of $t^*$ with $\mu(A) > 2 \cdot \sqrt{\epsilon_2} \cdot \mu(t^*)$. Then, 

$$\int_{(r, \theta^*) \in A} J_M(r, \theta^*) \cdot dr > \epsilon_2 \cdot 2 \cdot (1 - \cos(r_1))$$
\end{observation}

\noindent {\bf Proof:} Since $A$ is a measurable set, for every $\eta > 0$, there exists an elementary set $E_{\eta}$ 
such that $\mu((E_{\eta}-A) \cup (A-E_{\eta})) < \eta$. Note that an elementary set $E_{\eta}$ can be written as an almost disjoint
union of a finite number of intervals. In the following derivation, we assume that $\sqrt{\epsilon_2} < \frac{1}{2}$ and $r_1 < \frac{1}{2}$. Hence,

$$\int_{(r, \theta^*) \in E_{\eta}} J_M(r, \theta^*) \cdot dr \geq 2 \cdot \int_{0}^{\frac{\mu(E_{\eta})}{2}} \sin(x) dx = 2 \cdot \bigg( 1 - \cos \bigg(\frac{\mu(E_{\eta})}{2}\bigg) \bigg) $$

Further, since both $A - E_{\eta}$ and $E_{\eta} - A$ are measurable sets:

$$\int_{(r, \theta^*) \in A} J_M(r, \theta^*) \cdot dr = \int_{(r, \theta^*) \in E_{\eta}} J_M(r, \theta^*) \cdot dr + \int_{(r, \theta^*) \in A - E_{\eta}} J_M(r, \theta^*) \cdot dr - \int_{(r, \theta^*) \in E_{\eta}-A} J_M(r, \theta^*) \cdot dr$$

$$\geq \int_{(r, \theta^*) \in E_{\eta}} J_M(r, \theta^*) \cdot dr - \int_{(r, \theta^*) \in E_{\eta}-A} J_M(r, \theta^*) \cdot dr$$

Since $J_M(r, \theta^*) \leq \sin(r_1)$ for all values of $r$, $\int_{(r, \theta^*) \in E_{\eta}-A} J_M(r, \theta^*) \cdot dr \leq \mu(E_{\eta}-A) \cdot \sin(r_1) \leq \eta \cdot \sin(r_1)$. Thus, 

$$\int_{(r, \theta^*) \in A} J_M(r, \theta^*) \cdot dr \geq \int_{(r, \theta^*) \in E_{\eta}} J_M(r, \theta^*) \cdot dr - \eta \cdot \sin(r_1)$$

$$ \geq 2 \cdot \bigg( 1 - \cos \bigg(\frac{\mu(E_{\eta})}{2}\bigg) \bigg)   - \eta \cdot \sin(r_1)$$

Note that $\lim_{\eta \rightarrow 0} \mu(E_{\eta}) = \mu(A)$.
Taking the limit of the above inequality as $\eta \rightarrow 0$, we conclude that

$$\int_{(r, \theta^*) \in A} J_M(r, \theta^*) \cdot dr \geq  2 \cdot \bigg( 1 - \cos \bigg(\frac{\mu(A)}{2}\bigg) \bigg)$$

$$ \geq 2 \cdot \bigg( 1 - \cos \bigg( 2 \cdot \sqrt{\epsilon_2} \cdot r_1 \bigg) \bigg)$$
$$= \frac{1 - \cos \big( 2 \cdot \sqrt{\epsilon_2} \cdot r_1 \big)  }{1 - \cos(r_1) } \cdot \big( 2 \cdot (1 - \cos(r_1)) \big)$$

$$\geq \frac{ \frac{\big(2 \cdot \sqrt{\epsilon_2} \cdot r_1 \big)^2}{2} -  \frac{\big(2 \cdot \sqrt{\epsilon_2} \cdot r_1 \big)^4}{4!} }{ \frac{r_1^2}{2} } \cdot \big( 2 \cdot (1 - \cos(r_1)) \big)$$

$$= 4 \cdot \epsilon_2 \cdot \bigg( 1 - \frac{4 \cdot \epsilon_2 \cdot r_1^2}{12} \bigg) \cdot \big( 2 \cdot (1 - \cos(r_1)) \big)$$

$$ \geq 4 \cdot \epsilon_2 \cdot \bigg( 1 - \frac{4}{12} \bigg) \cdot \big( 2 \cdot (1 - \cos(r_1)) \big)$$

$$ > \epsilon_2 \cdot \big( 2 \cdot (1 - \cos(r_1)) \big)$$


\begin{observation}
The integral $\int_{r=-r_1}^{r_1} I_M(r, \theta^*) \cdot dr$ exists, and is at least $(1 - 2 \cdot \sqrt{\epsilon_2}) \cdot \mu(t^*)$. 
\end{observation}
    
\noindent {\bf Proof:} Note that, since $t^* \cap M$ is a measurable set, $\int_{r=-r_1}^{r_1} I_M(r, \theta^*) \cdot dr = \mu(t^* \cap M)$. Suppose the above statement is not true.
Thus, by Observation \ref{J-lower-bound},

$$\int_{(r, \theta^*) \in t^* - M} J_M(r, \theta^*) \cdot dr > \epsilon_2 \cdot 2 \cdot (1 - \cos(r_1))$$

Since $$\int_{(r,\theta^*) \in t^*} J_M(r, \theta^*) \cdot dr = 2 \cdot (1 - \cos(r_1))$$, we conclude that:

$$\int_{(r, \theta^*) \in t^* \cap M} J_M(r, \theta^*) \cdot dr < (1 - \epsilon_2) \cdot 2 \cdot (1 - \cos(r_1))$$

This contradicts Observation \ref{obs-tonelli-appl}, and hence the statement is true.
$\blacksquare$

 We conclude that:

\begin{corollary}
    $$\frac{\mu(t^* \cap M)}{\mu(t^*)} \geq 1 - 2 \cdot \sqrt{\epsilon_2} $$
\end{corollary}

Let $D_2$ be a disc with the same center as disc $D_1$, but with half the radius $r_2 = \frac{r_1}{2}$. Let $t_1 = t^* \cap D_2$.

\begin{observation}
\label{int-t1-M}
   $$\mu(t_1 \cap M) \geq (1 - 4 \cdot \sqrt{\epsilon_2}) \cdot \mu(t_1)$$
\end{observation}

\noindent {\bf Proof:} Note that:

    $$\mu(t_1 \cap M) = \mu(t_1) - \mu(t_1 - M)$$
    $$ \geq \mu(t_1) - \mu(t^* - M)$$
    $$ \geq \mu(t_1) - 2 \cdot \sqrt{\epsilon_2} \cdot \mu(t^*)$$
    $$ = (1 - 4 \cdot \sqrt{\epsilon_2}) \cdot \mu(t_1) ~(~since ~\mu(t^*) = 2 \cdot \mu(t_1) ~)$$

$\blacksquare$

\begin{observation}
\label{min-length-Gy-int-c2}
For every point $y_1 \in t_1$, 
    $$\mu( G(y_1) \cap c_2 ) \geq r_1$$
\end{observation}

\noindent {\bf Proof:} Let $y$ be the center of disc $D_2$. Let $y_1 \in t_1$. Let $L$ be
the lune of angle less than or equal to $\frac{\mu(t_1)}{2}$ formed by $G(y)$ and $G(y_1)$.
As one moves continuously from $y$ to $y_1$ along segment $t_1$, $G(y)$ continuously rotates to $G(y_1)$, inside the lune $L$.

In fact, each point $q$ of $G(y)$ moves continuously along a circular segment $g_q$ under this rotation. Further, the length of any circular segment $g_q$ is at most $\frac{\mu(t_1)}{2}$. 

Since $G(y)$ intersects $scaled(c_2)$, there exists a point $q’$ in $G(y) \cap c_2$ such that 

            $$d(q’, bd(c_2)) \geq r_1$$

(The above holds because $\mu(c_1)=\mu(c_2)$.)

Since length of $g_{q'}$ is at most $\frac{\mu(t_1)}{2}$, this implies that for any two points $u,v \in g_q$, $d(u,v) \leq \frac{\mu(t_1)}{2}$. Since $q' \in  scaled(c_2)$, we conclude that every point in $g_{q'}$ belongs to $c_2$.

Then, every point $z$ on the circular segment $g_{q’}$ has the following property:

        $$d(z, bd(c_2)) \geq d(q’, bd(c_2)) - d(q',z) ~(by ~triangle ~inequality)$$
        $$\geq r_1 -  \frac{\mu(t_1)}{2}$$
(since $q’$ belongs to $scaled(c_2)$, and $\mu(c_1) = \mu(c_2)$, $d(q’, bd(c_2)) \geq r_1$. )
        $$= r_1 - \frac{r_1}{2}$$
        $$= \frac{r_1}{2}$$

Let $w$ be any point of $t_1$. Let $w’$ be the point on the geodesic segment $g_{q'}$ corresponding to $w$. Since $g_{q'} \subset c_2$, $w' \in c_2$. Then,
$G(w) \cap c_2$ must contain a geodesic segment which extends till distance at least $d(w',bd(c_2))$ on both sides of $w'$.
This implies that, for every $w \in t_1$:

        $$\mu(G(w) \cap c_2) \geq 2 \cdot d(w’, bd(c_2)) \geq r_1$$

Hence the observation is proved $\blacksquare$

\begin{observation}
$$\mu(c_2 - M) \geq  2 \cdot \epsilon_1 \cdot \mu(c_2)$$
\end{observation}

\noindent {\bf Proof:} Define the set $U = \cup_{z \in t_1 \cap M} G(z)$. We now prove that $U$ is a measurable set. Let $\epsilon > 0$ be a real number. Since $t_1 \cap M$ 
is a measurable set, there exists an open set $O$ such that $t_1 \cap M \subset O$ and $\mu^*(O-(t_1 \cap M)) < \frac{\epsilon}{4}$. Suppose $O$ is the union of the countably infinite sequence of open intervals $s_1, s_2, s_3, \ldots$. Let $L(s_i)$ be the open lune formed by the set of points $\cup_{w \in s_i} G(w)$. Define $O'$ as the union of the countably infinite sequence of
open lunes $L(s_1), L(s_2), L(s_3), \ldots$. Then, $O'$ is an open set and $U \subset O'$.

Note that the measure of the open lune corresponding to an interval of length $l$ is 
$\frac{2 \cdot l}{2 \cdot \pi} \cdot 4 \cdot \pi = 4 \cdot l$. Thus, $\mu^*(O' - U) \leq 4 \cdot \mu^*(O-(t_1 \cap M)) < \epsilon$. Thus, for every $\epsilon > 0$, we can construct an open set $O'$ such that $U \subset O'$ and $\mu^*(O'-U) < \epsilon$. Hence, $U$ is a measurable set. Further, we conclude that $c_2 \cap U$ is also a measurable set. 

Since $M$ is orthogonal-pair-free, $c_2 \cap U \subset c_2 - M$. Let $I_{t_1}$ be the following indicator function: (i) $I_{t_1}(z) = 1$ if $z$ belongs to $G(y_1)$ for some point $y_1 \in t_1 \cap M$, and (ii) $I_{t_1}(z) = 0$ otherwise. Thus,

    $$\mu(c_2 - M) \geq \mu(c_2 \cap U) = \int_{z \in c_2} I_{t_1}(z) da$$

(Here $da$ denotes the area element of the 2-dimensional Lebesgue measure $\mu(\cdot)$ on the sphere.)

Let $L$ be the lune formed by the set of points $\cup_{z \in t_1} G(z)$. We chose a parametrization such that the two vertices of lune $L$ are the north and south poles. Thus, for every point $y_1 \in t_1 \cap M$, all points in $G(y_1) \cap c_2$ have the same value of parameter $\theta$. Further, we assume that the two endpoints of $t_1$ correspond to $\theta=0$ and $\theta=\mu(t_1)$ respectively. By an application of Tonelli's theorem \cite{royden}, we conclude that

    $$\mu(c_2 \cap U)= \int_{y_1 \in t_1 \cap M} \int_{(r,\theta) \in G(y_1) \cap c_2} |\sin(r)| \cdot dr \cdot d\theta$$


Let $U_M(\theta)=1$ if and only if the point $y \in t_1$ corresponding to 
angle $\theta$ belongs to $M$. Otherwise, $U_M(\theta)=0$. Thus,

$$\mu(c_2 - M) \geq \int_{\theta=0}^{\mu(t_1)} U_M(\theta) \cdot \bigg( \int_{(r,\theta) \in G(y_1) \cap c_2} |\sin(r)| \cdot dr \bigg) \cdot d\theta$$

By Observation \ref{obs-dist-from-lune-vertices}, for every point $w \in c_2$, $|\sin(r)| \geq \big| \sin \big( \frac{\pi}{8} \big) \big|$. Hence,

$$\mu(c_2 - M) \geq \int_{\theta=0}^{\mu(t_1)} U_M(\theta) \cdot \bigg( \int_{(r,\theta) \in G(y_1) \cap c_2} \sin \bigg( \frac{\pi}{8} \bigg) \cdot dr \bigg) \cdot d\theta$$

$$\geq \int_{\theta=0}^{\mu(t_1)} U_M(\theta) \cdot  \sin \bigg( \frac{\pi}{8} \bigg) \cdot r_1  \cdot d\theta ~(by ~Observation ~\ref{min-length-Gy-int-c2})$$

$$=   \sin \bigg( \frac{\pi}{8} \bigg) \cdot r_1 \cdot \mu(t_1 \cap M)$$

$$\geq  \sin \bigg( \frac{\pi}{8} \bigg) \cdot r_1 \cdot (1 - 4 \cdot \sqrt{\epsilon_2}) \cdot \mu(t_1) ~(by ~Observation ~\ref{int-t1-M})$$

$$= \sin \bigg( \frac{\pi}{8} \bigg) \cdot (1 - 4 \cdot \sqrt{\epsilon_2}) \cdot r_1^2$$

$$= \sin \bigg( \frac{\pi}{8} \bigg) \cdot (1 - 4 \cdot \sqrt{\epsilon_2}) \cdot \epsilon_1^{\frac{2}{N}} \cdot \mu(c_1)$$

$$= \sin \bigg( \frac{\pi}{8} \bigg) \cdot (1 - 4 \cdot \sqrt{\epsilon_2}) \cdot \epsilon_1^{\frac{2}{N}} \cdot \mu(c_2) ~(since ~\mu(c_1) = \mu(c_2))$$

$$= \sin \bigg( \frac{\pi}{8} \bigg) \cdot \bigg(1 - 4 \cdot \sqrt{\frac{32}{\pi} \cdot \epsilon_1^{1 - \frac{2}{N}}} \bigg) \cdot \epsilon_1^{\frac{2}{N}} \cdot \mu(c_2)$$

$$= \sin \bigg( \frac{\pi}{8} \bigg) \cdot \bigg(1 - \frac{16 \cdot \sqrt{2}}{\sqrt{\pi}} \cdot \epsilon_1^{1 - \frac{2}{N}} \bigg) \cdot \epsilon_1^{\frac{2}{N}} \cdot \mu(c_2)$$

By choice of constants (Section \ref{choice-of-constants}), this is at least $2 \cdot \epsilon_1 \cdot \mu(c_2)$, and hence the claim is proved $\blacksquare$

From the above observation, we conclude that:

    $$\frac{\mu(c_2 \cap M)}{\mu(c_2)} = 1 - \frac{\mu(c_2 - M)}{\mu(c_2)}$$
    
    $$\leq 1 - 2 \cdot {\epsilon_1}$$

We arrive at a contradiction, since the above quantity must be at least $1 - \epsilon_1$, and hence $M_2$ is orthogonal-pair-free $\blacksquare$

\subsection{Main result}

\begin{theorem}
\label{near-optimal-dyadic-cells}
$$\alpha_3 = \limsup_{S \in \mathcal{B}} \mu(S)$$
\end{theorem}

\noindent {\bf Proof:} Follow from Lemmas \ref{union-of-dense-cells} and \ref{scaled-orthogonal-pair-free} $\blacksquare$

\section{Near-optimality of the union of a finite number of mutually disjoint spherical convex sets}

\subsection{Geometric preliminaries}

For two distinct points $x,y \in \mathbb{S}^2$ such that $y \neq -x$, $\overline{xy}$ denotes the geodesic segment between $x$ and $y$ of length $d(x,y)$. The upper open hemisphere is a spherical convex set, and hence for any two distinct points $x,y \in H$, $\overline{xy} \subset H$.\\

\noindent {\it Polarity.} If great circle $G(a)$ passes through point $b$, then great circle $G(b)$ passes through point $a$.\\

\noindent {\it Pasch's axiom for the Euclidean plane.} Let $a,b,c$ be three distinct points in the Euclidean plane. If a line $l$ intersects the closed segment $\overline{ab}$, then it intersects $\overline{bc} \cup \overline{ca}$.\\

\noindent {\it Pasch's axiom for a triangle in upper open hemisphere.} Let $a,b,c$ be three distinct points in the upper open hemisphere $H$ of sphere $\mathbb{S}^2$. If a great circle $g$ intersects the closed geodesic segment $\overline{ab}$, then it intersects $\overline{bc} \cup \overline{ca}$.

We note that the Pasch's axiom for the upper open hemisphere can be derived from
the Pasch's axiom for Euclidean plane, by mapping the open hemisphere $H$ by gnomonic projection to
the tangent plane $T$ at the north pole.

\subsection{A new orthogonal-pair-free set}
Let $M \in \mathcal{B}$ be an orthogonal-pair-free set, consisting of a finite number of almost disjoint 
dyadic cells. Let $G(M) \subset \mathbb{S}^2$ denote the set of points: 

$$\{ z ~| ~z \in G(y) ~for ~some ~point ~y \in M\}$$ 

Define $M_1$ as the following set: 

$$M_1 = \{ z ~| ~G(z) \subset G(M) ~and ~z \in \mathbb{S}^2\}$$ 

Finally, define $M_2$ to be the set difference $M_1 – G(M)$. 

\begin{lemma} 
$M_2$ is orthogonal-pair-free and $M \subseteq M_2$. 
\end{lemma} 

\noindent {\bf Proof:} First, by definition, $M \subseteq M_1$. Thus, $M_2 = M_1 – G(M) \supseteq M – G(M) = M$. 
(Here we have used the set-theoretic fact: If $A \subseteq B$, $A-C \subseteq B-C$ for every set $C$.) 

Suppose, for the sake of contradiction, that $M_2$ is not orthogonal-pair-free. Let $y \in M_2$ be a point such that 
$G(y) \cap M_2 \neq \phi$. Since $G(y) \subset G(M)$, this implies that $G(M) \cap M_2 \neq \phi$. We arrive at 
a contradiction, since $M_2 = M_1 – G(M)$ $\blacksquare$

\subsection{Triangle Lemma}

We first prove a basic lemma:

\begin{lemma}
Let $x$, $y$, and $z$ be three distinct points such that $\overline{xy} \subset M_2$ and $\overline{yz} \subset M_2$.
Then, $\overline{xz} \subset M_1$.
\end{lemma}

\noindent {\bf Proof:} Since $M_2$ is orthogonal-pair-free, $d(x,y) < \frac{\pi}{2}$ and $d(y,z) < \frac{\pi}{2}$. 
Let $H$ be an open hemisphere with north pole at $y$. Then, $H$ contains all three points $x$, $y$, and $z$.

Since $\overline{xy} \cup \overline{yz} \subseteq M_2 \subseteq M_1$, 

$$\cup_{a \in \overline{xy} \cup \overline{yz}} G(a) \subseteq G(M)$$

We complete the proof by showing that:

$$\cup_{a \in \overline{xz}} G(a) \subseteq \cup_{a \in \overline{xy} \cup \overline{yz}} G(a)$$

Let $w \in \cup_{a \in \overline{xz}} G(a)$. By polarity, $G(w) \cap \overline{xz} \neq \phi$. By Pasch's axiom applied on $\triangle xyz$ in the upper open hemisphere $H$, $G(w) \cap \big( \overline{xy} \cup \overline{yz} \big) \neq \phi$. This implies by polarity that $w \in \cup_{a \in \overline{xy} \cup \overline{yz}} G(a)$ $\blacksquare$

\begin{lemma} 
(Triangle lemma) Let $x, y, z$ be three distinct points in $M_2$. Suppose $\overline{xy} \subset M_2$ and 
$\overline{yz} \subset M_2$. Then, 

\begin{enumerate}
\item $\overline{xz} \subset M_2$. 
\item $d(x,y) < \frac{\pi}{2}$, $d(y,z) < \frac{\pi}{2}$, and $d(x,z) < \frac{\pi}{2}$.
\end{enumerate}
\end{lemma} 

\noindent {\bf Proof:} Since $M_2$ is orthogonal-pair-free, 
$d(x,y) < \frac{\pi}{2}$ and $d(y,z) < \frac{\pi}{2}$. Rotate
the sphere so that $y$ goes to the north pole. Then, $\triangle xyz$ lies completely in the upper open hemisphere.

By the above lemma, $\overline{xz} \subset M_1$. Suppose, for the sake of contradiction, that $\overline{xz}$ is not a subset of $M_2$. Since $M_2 = M_1 - G(M)$, this implies that there exists a point $w \in M$
such that $G(w)$ intersects $\overline{xz}$. There are the following three cases:

\begin{enumerate} 

\item {\it $G(w)$ passes through $z$}. In this case, $d(z,w) = \frac{\pi}{2}$. Since $M \subseteq M_2$, $w \in M_2$, and this contradicts our assumption
that $M_2$ was orthogonal-pair-free.

\item {\it $G(w)$ passes through $x$}. Similar to Case I above.

\item {\it $G(w)$ passes through an interior point of segment $\overline{xz}$}. By the 
Pasch's axiom for the upper open hemisphere applied on the spherical triangle $\triangle xyz$, $G(w)$ must pass through a 
point $w'$ of the set $\overline{xy} \cup \overline{yz}$. Since 
$\overline{xy} \cup \overline{yz} \subseteq M_2$, $w \in M \subset M_2$ and $d(w, w') = \frac{\pi}{2}$, this contradicts our assumption that
$M_2$ was orthogonal-pair-free.
\end{enumerate}

This proves the first part of the lemma. We now prove the second part of the lemma.
Since $M_2$ is orthogonal-pair-free, $d(x,y) < \frac{\pi}{2}$, $d(y,z) < \frac{\pi}{2}$,
and $d(x,z) < \frac{\pi}{2}$.

Hence the lemma is proved $\blacksquare$

\subsection{Piecewise-linear lemma}
A finite piecewise linear path is a path consisting of a finite number of geodesic
segments. For any two points $x, y \in M_2$, we say that $x \sim_{M_2} y$ if and only if there exists a finite piecewise linear path $P_{xy}$ from $x$ to $y$ on the surface of the sphere $\mathbb{S}^2$ such that every point 
$z \in P_{xy}$ belongs to the set $M_2$.

\begin{lemma} 
(Piecewise-linear lemma) Let $x, y \in M_2$. If $x \sim_{M_2} y$, then $\overline{xy} \subset M_2$. 
\end{lemma} 

\noindent {\bf Proof:} By the definition of $x \sim_{M_2} y$, there exists a piecewise linear path from 
$P_{xy} \subseteq M_2$ from $x$ to $y$ on the sphere, such that $P_{xy}$ consists of a finite number of geodesic segments. 

Suppose $P_{xy}$ consists of $m$ geodesic segments $\overline{\rm s_0s_1}$, $\overline{\rm s_1s_2}$, $\ldots$, 
$\overline{\rm s_{m-1}s_m}$, where $s_0=x$ and $s_m=z$. 

Applying the triangle lemma to $s_0, s_1, s_2$, we conclude that $\overline{s_0 s_2} \subset M_2$. Since $M_2$ is 
orthogonal-pair-free, we also conclude that length of $\overline{s_0 s_2}$ is less than $\frac{\pi}{2}$. 

Applying the triangle lemma to $s_0, s_2, s_3$, we conclude that $\overline{s_0 s_3} \subset M_2$. Since $M_2$ is 
orthogonal-pair-free, we also conclude that length of $\overline{s_0 s_3} < \frac{\pi}{2}$. 

Proceeding in this manner, we finally conclude that $\overline{s_0 s_m} \subset M_2$ and length of this segment is less than 
$\frac{\pi}{2}$. 

Hence the lemma is proved $\blacksquare$

We have the following corollary: 

\begin{corollary} 
\label{piecewise-linear-corollary}
$\sim_{M_2}$ is an equivalence relation. 
\end{corollary} 

\noindent {\bf Proof:} By definition, $\sim_{M_2}$ is reflexive and symmetric. (We assume that a single point 
is a piecewise-linear path of length $0$.) 

We now prove that $\sim_{M_2}$ is transitive. Let $x,y,z$ be three distinct points such that $x \sim_{M_2} y$ 
and $y \sim_{M_2} z$. Suppose $P_{xy} \subset M_2$ and $P_{yz} \subset M_2$ are finite piecewise-linear paths from $x$ to $y$ 
and from $y$ to $z$, respectively. 

Let $Q$ be the concatenation of these two paths. $Q$ is finite piecewise-linear. 
Hence $x \sim_{M_2} z$, and $\sim_{M_2}$ is transitive.

If $x=y$, $y=z$ or $x=z$, transitivity follows from definitions. 

We thus conclude that $\sim_{M_2}$ is an equivalence relation $\blacksquare$ 

For a point $x \in M_2$, let $[x]_{M_2}$ denote the equivalence class of $x$ under relation $\sim_{M_2}$. 
Recall that if $[x]_{M_2} \cap [y]_{M_2} \neq \phi$, then $[x]_{M_2}=[y]_{M_2}$. 

\begin{lemma} 
Let $x \in M_2$. Then $[x]_{M_2}$ is a spherical convex set, and $[x]_{M_2} \subset M_2$. 
\end{lemma} 

\noindent {\bf Proof:} Suppose $y_1, y_2 \in [x]_{M_2}$ be two distinct points. Then $y_1 \sim_{M_2} y_2$, and by the 
piecewise-linear lemma $\overline {y_1y_2} \subset M_2$. 

Let $P_{xy_1}$ be a piecewise-linear path from $x$ to $y_1$. Let $w$ be any point on the segment $\overline {y_1y_2}$. 
Then $Q= P_{xy_1} \circ \overline {y_1 w}$ is a piecewise-linear path from $x$ to $w$. Since 
$\overline{y_1 y_2} \subset M_2$, this implies that $Q \subset M_2$. Hence, we conclude that $x \sim_{M_2} w$, and 
hence $w \in [x]_{M_2}$. 

We conclude that $\overline{y_1y_2} \in [x]_{M_2}$, and hence the lemma is proved $\blacksquare$

\subsection{First convexification operation $conv_1(\cdot)$}
Suppose $M$ consists of $w$ almost disjoint dyadic cells. Now, we define the {\it first convexification} $conv_1(M)$ of set $M$ as the collection $\{ [x]_{M_2} ~| ~x \in M\}$ of equivalence classes, after removing duplicate entries.

\begin{lemma} 
\label{conv1-properties}
\begin{enumerate} 
\item $conv_1(M)$ is a finite union of at most $w$ spherical convex sets, where $w$ is the number of almost 
disjoint dyadic cells in $M$. 

\item $conv_1(M)$ is orthogonal-pair-free. 

\item $conv_1(M)$ is a measurable set, and since $M \subseteq conv_1(M)$: 

                          $$\mu(M) \leq \mu(conv_1(M))$$ 
\end{enumerate} 
\end{lemma} 

\noindent {\bf Proof:} Note that $conv_1(M) \supseteq M$. Let $c$ be a dyadic cell of $M$. Then, for any two points 
$z_1, z_2 \in c$, there exists a piecewise-linear path from $z_1$ to $z_2$ inside cell $c$. Since $M \subset conv_1(M)$, 
this implies that all points of $c$ belong to the same equivalence class under $\sim_{M_2}$. Thus, $conv_1(M)$ is a union 
of at most $w$ distinct equivalence classes. By the above lemma, each equivalence class is a spherical convex set, and \
further any two equivalence classes are disjoint. 

Since $conv_1(M) \subset M_2$, this implies that $conv_1(M)$ is also orthogonal-pair-free. 

Finally, every convex spherical set is measurable, and hence a finite union of disjoint spherical 
convex sets is also measurable (by finite additivity of measure). Hence, since $conv_1(M) \supseteq M$, 
$\mu(conv_1(M)) \geq \mu(M)$ $\blacksquare$

\subsection{Second convexification operation $conv_2(\cdot)$}
\noindent {\it Some geometric facts.} The geometric facts used here for spherical convex sets follow from the corresponding results for Euclidean convex sets \cite{kelly-weiss}, as the two are related by gnomonic projection. 

The distance $d(A,B)$ between two sets $A$ and $B$ is equal to $inf_{x 
\in A, y \in B} d(x,y)$.

If $A$ is a convex set, $cl(A)$ is also a convex set. Further, $\mu(A) = \mu(int(A)) = \mu(cl(A))$. Let $A$ and $B$ be two convex sets on the sphere. If $d(A,B) = 0$, $bd(A) \cap bd(B) \neq \phi$. 

Let $A$ be a convex set with a non-empty interior. If $x,y \in int(A)$, $\overline{xy} \subset int(A)$. Let $x \in int(A)$ and $y \in bd(A)$, then there exists a piecewise-linear path $P \subset cl(A)$ with $2$ geodesic segments $\overline{xz}$ and $\overline{zy}$ such that $z \in int(A)$ and $\overline{zy} \cap bd(A) = \{y\}$.

Let $N_1 = conv_1(M)$ be the set obtained by applying the first convexification
operation to set $M$. Suppose $N_1$ consists of $w$ disjoint spherical convex sets
$D_1, D_2, \ldots, D_w$.

\begin{observation}
$$\mu(int(N_1)) = \mu(N_1)$$
\end{observation}

\noindent {\bf Proof:} First observe that $int(N_1) = \cup_{i=1}^{w} int(D_i)$.
Since each $D_i$ is a spherical convex set, we have that $\mu(int(D_i)) = \mu(D_i)$. Further, since $D_i \cap D_j = \phi$ for all $1 \leq i < j \leq w$, we have that:

$$\mu(int(N_1)) = \sum_{i=1}^{w} \mu(int(D_i)) = \sum_{i=1}^{w} \mu(D_i) = \mu(N_1)$$

$\blacksquare$

Note that though $N_1$ consists of $w$ mutually disjoint spherical convex sets, it is possible
that for some $1 \leq i < j \leq w$, $d(D_i, D_j)=0$. 

We now describe a {\it second convexification} operation $conv_2(N_1)$, which returns an orthogonal-pair-free set $N_2$ such that (i) $\mu(N_2) \geq \mu(N_1)$, (ii) $N_2$ consists of $w'$ spherical convex components $E_1, E_2, \ldots, E_{w'}$ where $1 \leq w' \leq w$, and (iii) $\min_{1 \leq i < j \leq w'} d(E_i, E_j) > 0$.

$conv_2(\cdot)$ applies the $conv_1(\cdot)$ operation repeatedly, as per the following algorithm:

\begin{enumerate}
    \item Set $N_2=N_1$.
    
	\item If every two spherical convex sets in $N_2$ are at positive distance, halt.

	\item Else, let $E_i$ and $E_j$ ($i \neq j$) be two disjoint spherical convex sets in $N_2$ at distance $0$ from each other.

		\begin{enumerate}
			\item Let $p$ be a point in the intersection of $bd(E_i)$ and $bd(E_j)$.
			\item Define $N' = int(N_2) \cup \{ p \}$.
			\item Define $N'' = conv_1(N')$.
			\item Set $N_2=N''$ and go to Step $2$ above.
		\end{enumerate}
\end{enumerate}

\begin{observation}
Consider an execution of Step $3$ of the above algorithm. Suppose $d(E_i, E_j)=0$. Then there exists a point $p \in bd(E_i) \cap bd(E_j)$.
\end{observation}

\noindent {\bf Proof:} Follows from the geometry of spherical convex sets $\blacksquare$

\begin{lemma}
Consider an execution of Step $3$ of the above algorithm.
Then, $N' = int(N_2) \cup \{ p \}$ is orthogonal-pair-free.
\end{lemma}

\noindent {\bf Proof:} Suppose, for the sake of contradiction, that $int(N_2) \cup \{ p \}$ is not orthogonal-pair-free.
Then, there exists a point $z \in int(N_2)$ such that $G(z)$ passes through $p$. 

Since $z \in int(N_2)$, there exists an $\epsilon>0$ such that $B(z, \epsilon) \subset int(N_2)$. Then the set $S_{z, \epsilon} = \cup_{q \in B(z, \epsilon)} G(y)$ is a spherical strip of width $\epsilon$ with $G(z)$ as its middle great
circle.

Since $G(z)$ passes through $bd(E_i) \cap bd(E_j)$, the spherical strip $S_{z,\epsilon}$ contains
a point $z' \in int(E_i) \cup int(E_j) \subset int(N_2)$. Thus, $d(z,z') = \frac{\pi}{2}$, 
where both $z,z' \in int(N_2)$. This contradicts our assumption that $int(N_2)$ was orthogonal-pair-free $\blacksquare$

\begin{observation}
Suppose Step $3$ of the above algorithm is executed. Then, $N''$ is orthogonal-pair-free.
\end{observation}

\noindent {\bf Proof:} Follows from Theorem \ref{conv1-properties} $\blacksquare$

\begin{lemma}
Every time Step $2$ of the above algorithm is executed, the number of spherical convex sets in $N_2$ decreases by at least $1$.
\end{lemma}

\noindent {\bf Proof:} Let $x \in int(E_i)$ and $y \in int(E_j)$. There is a piecewise-linear path from $x$ to $p$ in 
$int(E_i) \cup \{ p \}$. Similarly, there is a piecewise-linear
path from $p$ to $y$ in $int(E_j) \cup \{ p \}$. Thus,
 there is a finite piecewise-linear path between $x$ and $y$ in set $N''$. Further, there is geodesic segment in $N''$ between any two points $x,y$ in $int(E_i)$ as well as between any two points $x,y \in int(E_j)$. Thus, by Corollary \ref{piecewise-linear-corollary}, all points in $int(E_i) \cup int(E_j) \cup \{ p \}$ belong to the same equivalence class $\blacksquare$

\begin{lemma}
\label{second-convexify}
Let $N_2$ be the final orthogonal-pair-free set obtained after executing the above algorithm. Then $N_2$ consists of at most $w$ mutually disjoint convex sets, such that the distance between any two convex sets is strictly greater than $0$. 
\end{lemma}

\noindent {\bf Proof:} The algorithm halts when the number of convex sets reduces to $1$ (up to antipodal symmetry), or every two convex sets are at positive distance. $\blacksquare$

\subsection{Final convexification operation}

We define the {\it final convexification} $conv(M)$ of $M$ as $conv_2(conv_1(M))$ i.e.,
as the the composition of the first and second convexification operations defined above.

\begin{theorem}
\label{final-convexify}
For every $M \in \mathcal{B}$, $conv(M) \in \mathcal{A}$ and $\mu(conv(M)) \geq \mu(M)$.
\end{theorem}

\noindent {\bf Proof:} From Lemmas \ref{conv1-properties} and \ref{second-convexify} $\blacksquare$

\section{Existence of an optimal solution among unions of at most $k$ mutually disjoint spherical convex sets}

\subsection{Preliminaries: Blaschke selection theorem}

For two spherical convex sets $U$ and $V$, their Hausdorff distance $\delta_H(U,V) = \max \big( \max_{x \in U} d(x,V), \max_{y \in V} d(U, y) \big)$. An equivalent definition of Hausdorff distance is the following. Let $B(r)$ denote the disc of radius $r$ and $+$ denote Minkowski sum. Let $\epsilon_1 > 0$ be the smallest real number such that $V \subset U + B(\epsilon)$ for all $\epsilon > \epsilon_1$. Let $\epsilon_2 > 0$ be the smallest real
number such that $U \subset V + B(\epsilon)$ for all $\epsilon > \epsilon_2$. Then, $\delta_H(U,V) = \min ( \epsilon_1, \epsilon_2 )$.

\begin{theorem}
(Blaschke selection theorem \cite{schneider}) Let $U_1, U_2, \ldots$ be a sequence of
spherical convex sets. Then, there exists a subsequence $U_{i_1}, U_{i_2}, \ldots$ ($1 \leq i_1 < i_2 < \cdots$) and
a spherical convex set $U^*$ such that $\lim_{j \rightarrow \infty} \delta_H(U_{i_j}, U^*) = 0$.
\end{theorem}

\subsection{Existence of limit set}

\begin{lemma} 
Let $M_1, M_2, \ldots$ be a countably infinite sequence of orthogonal-pair-free sets, such that each $M_i \in \mathcal{A}_k$ and $\limsup_{i \in \mathbb{N}} \mu(M_i) = \limsup_{S \in \mathcal{A}_k} \mu(S)$. Then, there exists a countably infinite
subsequence $M_{d_1}, M_{d_2}, \ldots$ of this sequence and a set $M^*$ such that the subsequence converges to $M^*$ in Hausdorff distance i.e., $\lim_{j \rightarrow \infty} \delta_H(M_{d_j}, M^*) = 0$. 
\end{lemma}

\noindent {\bf Proof:} We assume that each $M_i$ has exactly $k$ spherical convex sets (up to antipodal symmetry). If this is not the case, by pigeonhole principle, there exists an integer $k'$ ($1 \leq k' \leq k$) and a countably infinite 
subsequence of $M_1, M_2, \ldots$, with each set having exactly $k'$ spherical convex sets. The following argument can then be applied to this subsequence.

For $j \in \mathbb{N}$, let $C^{1}_{j}, C^{2}_{j}, \ldots, C^{k}_{j}$ be the disjoint spherical convex sets of $M_{j}$. We will repeatedly apply Blaschke selection theorem at most $k$ times. First, consider the sequence $C^1_{1}, C^1_{2}, \ldots$. Since this is a countably infinite sequence of spherical convex sets, by Blaschke selection theorem, there exists a countably infinite subsequence $C^1_{a_{1,1}}, C^1_{a_{1,2}}, \ldots$ 
, which converges in Hausdorff distance to a spherical convex set $C^*_1$. 
Next consider the countably infinite subsequence $C^2_{a_{1,1}}, C^2_{a_{1,2}}, \ldots$. Again, by Blaschke selection theorem, there exists a countably infinite subsequence $C^2_{a_{2,1}}, C^2_{a_{2,2}}, \ldots$, which converges in Hausdorff distance to a spherical convex set $C^*_2$. 

Applying the above argument at most $k$ times, we get a countably infinite subsequence 
$M_{a_{k,1}}, M_{a_{k,2}}, \ldots$ such that $\lim_{j \rightarrow \infty} \delta_H(C^{i}_{a_{k,j}}, C^*_i) = 0$ for every $1 \leq i \leq k$. Since there are a finite number of convex spherical sets, this leads to the conclusion that $\lim_{j \rightarrow \infty} \max_{1 \leq i \leq k} \big( \delta_H(C^{i}_{a_{k,j}}, C^*_i) \big) = 0$.

Thus, for every $\epsilon > 0$, there exists a natural number $N$ such that, for each $1 \leq i \leq k$ and $j > N$,

 $$C^i_{a_{k,j}} + B(\epsilon) \supset C^*_i$$ 
 $$C^*_i + B(\epsilon) \supset C^i_{a_{k,j}}$$   

Take $d_j = a_{k,j}$ for $j \in \mathbb{N}$ and $M^*=(C^*_1, C^*_2, \ldots, C^*_k)$. We conclude that $\lim_{j \rightarrow \infty} \delta_H(M_{d_j}, M^*) = 0$. 
$\blacksquare$ 

{\it Remark 1.} Note that, although $M^*$ can be described using a finite number of convex sets, it may not belong to $\mathcal{A}_k$. We next prove some properties of $\cup_{i=1}^{k} int(C^*_i)$, and then apply the convexification operation to obtain a set in $\mathcal{A}_k$.

\subsection{Interior of limit set is orthogonal-pair-free}

\begin{observation} 
Let $y \in \cup_{i=1}^{k} int(C^*_i)$. Let $\epsilon > 0$ be a real number such that $B(y,\epsilon) \subset \cup_{i=1}^{k} int(C^*_i)$. 
Then, there exists a $N_y > 0$ such that $B(y, \frac{\epsilon}{2}) \subset M_{d_w}$ for every 
$w > N_y$. 
\end{observation} 

 \noindent {\bf Proof:} Without loss of generality, assume that $B(y, \epsilon) \subset int(C^*_1)$. 
Take $N_y$ to be the smallest integer such that $\delta_H(C^1_{d_j}, C^*_1) < \frac{\epsilon}{4}$ 
for every $j > N_y$. Due to convergence in Hausdorff distance proved above, such a $N_y$ will exist. 

Now, suppose for the sake of contradiction, that there exists a $j' > N_y$ such that $C^1_{d_{j'}}$ 
does not contain $B(y, \frac{\epsilon}{2})$. Then, there exists a point $y_1 \in B(y, \frac{\epsilon}{2})$ such 
that $y_1 \notin C^1_{d_{j'}}$. Since $C^1_{d_{j'}}$ is a closed convex set, by the separating hyperplane theorem \cite{kelly-weiss}, there exists a great circle 
$g$ such that $g$ separates $int(C^1_{d_{j'}})$ from $y_1$. Thus, there exists a point $y_2$ on the larger geodesic disc $B(y, \epsilon)$ such that $d(y_2, g) \geq \frac{\epsilon}{2}$. We conclude that 
$d(y_2, C^1_{d_{j'}}) \geq d(y_2, g) \geq \frac{\epsilon}{2}$. This implies that $C^1_{d_{j'}} + \frac{\epsilon}{4}$ is not a superset of $C^*_1$.  
This contradicts our assumption that $\delta_H(C^1_{j}, C^*_1) < \frac{\epsilon}{4}$ for all $j > N_y$ $\blacksquare$

\begin{lemma} 
For $1 \leq i_1 < i_2 \leq k$, $int(C^*_{i_1}) \cap int(C^*_{i_2}) = \phi$. 
\end{lemma} 

\noindent {\bf Proof:} Suppose this is not the case. Then there exists a point $y$ and two integers $i_1, i_2$ ($1 \leq i_1 < i_2 \leq k$) such that the interiors of both $C^*_{i_1}$ and $C^*_{i_2}$ contain 
$y$. Thus, there exist $\epsilon_1>0$ and $\epsilon_2 > 0$ such that $B(y, \epsilon_1) \subset 
int(C^*_{i_1})$ and $B(y, \epsilon_2) \subset int(C^*_{i_2})$. Taking $\epsilon=\min(\epsilon_1, \epsilon_2)$, 
$B(y, \epsilon)$ is contained in both $int(C^*_{i_1})$ and $int(C^*_{i_2})$. 

By two applications of the above observation, we conclude that there exists a $N > 0$ such that 
$B(y, \frac{\epsilon}{2}) \subset C^1_{d_j}$ and $B(y, \frac{\epsilon}{2}) \subset C^2_{d_j}$, 
for every $j > N$. 

This implies that $int(C^1_{d_j}) \cap int(C^2_{d_j}) \neq \phi$ for every $j_w > N$. This contradicts 
our initial assumption that $M_{j_w}$ is finite union of mutually disjoint spherical convex sets (by disjoint, 
we mean that any two spherical convex sets are at positive distance from each other) $\blacksquare$ 

\begin{lemma} 
The $\cup_{i=1}^{k} int(C^*_i)$ is orthogonal-pair free. 
\end{lemma} 

\noindent {\bf Proof:} Suppose this is not the case. Then, there exist points $y, y_1 \in \cup_{i=1}^{k} int(C^*_i)$ such that $y_1 \in G(y)$. Suppose, without loss of generality (up to a relabeling), 
that $y \in int(C^*_1)$ and $y_1 \in int(C^*_2)$. 

Suppose $B(y, \epsilon_1) \subset int(C^*_1)$ and $B(y_1, \epsilon_2) \subset int(C^*_2)$, for 
some $\epsilon_1, \epsilon_2 > 0$. Then, by applying the above observation, there exists $N_1 > 0$ such that 
$B(y, \frac{\epsilon_1}{2}) \subset C^1_{d_w}$ for every $w > N_1$. Similarly, there exists $N_2 > 0$ such that 
$B(y_1, \frac{\epsilon_2}{2}) \subset C^2_{d_w}$ for every $w > N_2$. Take $N=\max(N_1, N_2)$. Then, both $y$ and $y_1$ 
belong to $int(M_{d_w})$ for every $w > N$. 
This contradict our initial assumption that $int(M_{d_w})$ was orthogonal-pair-free $\blacksquare$

\subsection{Convexification of the limit set}

Let $M^{**} = conv( \cup_{i=1}^{k} int(C^*_i) ) $. By Theorem \ref{final-convexify}, we conclude that
(i) $\mu(M^{**}) \geq \mu(M^*)$ and (ii) $M^{**} \in \mathcal{A}_k$.
Hence $M^{**}$ is the optimal set in $\mathcal{A}_k$, and Theorem \ref{finite-k} is proved.

\end{document}